\documentclass[]{interact}

\usepackage{epstopdf}

\usepackage[natbibapa,nodoi]{apacite}
\usepackage[utf8]{inputenc}
\usepackage{tikz}
\usepackage{booktabs}
\usetikzlibrary{arrows.meta, positioning, calc, shapes.geometric}
\usepackage{subcaption}  

\usepackage{algpseudocode}
\usepackage{algorithmicx}
\usepackage{algcompatible}
\usepackage{algorithm}
\usepackage{amssymb}
\usepackage{bbm}
\usepackage{amsmath}
\usepackage{xcolor}
\usepackage{caption}
\usepackage{url} 
\usepackage{ifthen}
\newcommand{\Indent}{\hspace{1em}}

\begin{document}

\articletype{ARTICLE TEMPLATE}

\title{Anti-Windup in PID Control: Review, Analysis, and New Tuning Directions}

\author{
\name{M. Caparroz\textsuperscript{a}\thanks{CONTACT M. Caparroz. Email: mcaparroz@ual.es} and K. Soltesz\textsuperscript{b} and T. H\"agglund\textsuperscript{b} and J.~L. Guzm\'an\textsuperscript{a}}
\affil{\textsuperscript{a}Dep. of Informatics, Universidad de Almer\'ia, ceiA3, CIESOL, Ctra. Sacramento s/n, 04120 Almer\'ia, Spain; \textsuperscript{b}Dep. of Automatic Control, Lund University, Box 118 SE-22100 Lund, Sweden}
}

\maketitle 

\begin{abstract}
Actuator saturation is a fundamental nonlinearity that significantly degrades the performance of PID-controlled systems by inducing integrator windup, leading to overshoot, slow recovery, and even instability. Although numerous anti-windup strategies have been proposed, their practical tuning remains largely heuristic and suboptimal in many industrial scenarios. This paper presents a comprehensive comparative study of classical and advanced anti-windup techniques for PI-controlled first-order-plus-dead-time (FOPDT) processes under a wide range of operating conditions. The analysis includes dynamic and instantaneous back-calculation, conditional integration, and adapted schemes. In addition, a novel hybrid anti-windup strategy is proposed, combining conditional integration with dynamic back-calculation to improve responsiveness during saturation, whilst preserving smooth recovery dynamics. Moreover, a key contribution of this work is the development of systematic tuning rules for the tracking time constant in back-calculation schemes, specifically optimised for load-disturbance rejection. These rules are derived from an extensive optimisation study that considers the saturation ratio, controller aggressiveness, and disturbance characteristics. The resulting guidelines provide simple yet effective formulas that achieve near-optimal performance without requiring complex computations. Simulation results demonstrate that the proposed methods significantly outperform commonly used heuristic rules, particularly in disturbance rejection scenarios, and provide clear, practical recommendations for selecting and tuning anti-windup strategies in industrial applications.
\end{abstract}





\section{Introduction}

Proportional-Integral-Derivative (PID) controllers are the cornerstone of industrial automation, maintaining their status as the most widely used control strategy due to their inherent simplicity, robustness, and effectiveness across diverse process dynamics \citep{astrom_advanced_2006,visioli_practical_2006,rojas2021industrial,willis1999proportional}. Despite the rise of advanced control techniques, the PID's ease of implementation and the industry's familiarity with its three-term structure ensure its continued dominance. However, the practical application of PID control is frequently challenged by physical constraints, most notably actuator saturation \citep{zaccarian2011modern,hui1997new}.

Actuator saturation is a critical nonlinearity that degrades the performance of feedback control systems by inducing integrator windup and prolonging recovery from disturbances \citep{kothare1997stability}. When the controller’s output reaches the physical limits of an actuator, the feedback loop is rendered inactive, and the system enters an open-loop state: the actuator can no longer respond to further increases in the control signal. During this period, the integral term continues to accumulate error, resulting in windup. This accumulation can lead to significant performance degradation once the system returns to the linear region, characterised by excessive overshoots, prolonged recovery times, and, in severe cases, closed-loop instability \citep{astrom_advanced_2006,kothare1994unified,aastrom2021feedback}. 

To mitigate these effects, various anti-windup strategies have been developed in the literature to maintain controller performance and integrity during saturation. Classical approaches such as back-calculation and integrator clamping remain widely used in industry due to their simplicity and low computational cost \citep{astrom_advanced_2006,okelola2020performance}, whilst more advanced methods extend these ideas to more complex controller structures and architectures. 

Despite this extensive body of work, selecting an appropriate anti-windup strategy and tuning its parameters remains a challenging and often ad hoc task in practice. A large number of anti-windup techniques have been proposed over the years, yet their practical implementation and tuning still lack systematic, widely accepted guidelines \citep{okelola2020performance,sarhadi2026simple,yapp2024new,amiri2024anti}. The performance of a given technique depends not only on the controller design and process dynamics but also on the specific control objective, such as setpoint tracking or disturbance rejection, and on the nature of the disturbances involved \citep{cao2004anti,wang2016control,torstensson2013comparison}. As a result, a method that performs well in one scenario may lead to significant performance degradation in another, particularly in the presence of uncertainties, nonlinearities, and varying operating conditions.

In particular, even within a single class of methods such as back-calculation, the choice of the tracking time constant plays a critical role in determining the trade-off between fast recovery from saturation and closed-loop stability. Although simple heuristic rules, such as setting the tracking time equal to the integral time, are commonly used in practice, they often fail to provide satisfactory performance across different operating situations \citep{rundqwist_anti-reset_1990,buusek2017iae,markaroglu_tracking_2006,kumar2012comparative}.

This lack of systematic guidelines creates a gap between the wide availability of anti-windup techniques and their effective application in real systems. Practitioners frequently face multiple design choices: selecting the anti-windup structure, tuning its parameters, and adapting to different types of control problems, without clear criteria to guide these decisions.

In this context, this paper aims to provide a comprehensive and practical framework for selecting and tuning anti-windup strategies. First, several classical and recently proposed techniques are reviewed and compared across a wide range of operating conditions, including setpoint tracking and disturbance rejection, with varying process dynamics and saturation levels. Then, a new hybrid anti-windup scheme is proposed that combines back-calculation and conditional integration to improve performance during saturation while preserving smooth recovery dynamics.

Furthermore, a set of systematic tuning rules for the tracking time constant in back-calculation schemes is developed. The proposed rules are derived from an optimisation procedure conducted across a wide range of operating conditions, characterised by the saturation ratio, controller aggressiveness, and the disturbance-to-process-dynamics ratio, and are validated through simulation studies of first-order-plus-dead-time processes with PI controllers. The resulting guidelines allow practitioners to achieve near-optimal performance while maintaining implementation simplicity.

The article is structured as follows: Section \ref{sec: windup} describes the windup problem and its effect on the process output in the different control problems one may encounter. Section \ref{sec: review} presents a review of already existing solutions, in which both classical and recently proposed algorithms are exposed. Section \ref{sec: compared solutions} describes in depth the solutions that will be compared later on. The schemes and algorithms are presented, along with their main advantages and disadvantages. Section \ref{sec: New solutions} describes the new proposed solutions and the main contribution of the paper. This includes a scheme that combines back-calculation and conditional integration, and new tuning rules for $T_t$ when back-calculation is used. All the exposed rules are then compared in Section \ref{sec: validation examples}, where a wide variety of problems are analysed, culminating in practical rules that help the reader choose the best anti-windup solution for their particular problem. Finally, some conclusions are presented in Section \ref{sec: conclusions}. 

\section{The windup problem} \label{sec: windup}

\begin{figure}[h!]
        \centering
        \includegraphics[width=0.8\textwidth]{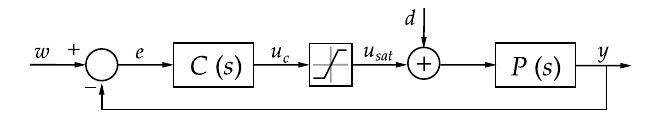}
        \caption{General feedback scheme for setpoint tracking and disturbance rejection}
        \label{fig:disturbance rejection esquema}
\end{figure}

Actuator saturation introduces a fundamental nonlinearity into feedback control systems, altering the nominal closed-loop behaviour assumed during controller design. Figure \ref{fig:disturbance rejection esquema} illustrates the standard feedback configuration for setpoint tracking and disturbance rejection problems, where $w$ is the setpoint for the controlled variable $y$, $e$ is the error between $w$ and $y$, $d$ is the load disturbance, $u_c$ is the controller output, $u_{sat}$ is the applied control action, after saturation, and $P(s)$ and $C(s)$ are the process and the PID controller, respectively, in Laplace domain.


Under nominal (unsaturated) conditions, the controller computes a control signal, $u_c$, according to:
\begin{equation}
    C(s) = \frac{u_c(s)}{e(s)} = K_p +\frac{K_i}{s} + K_d s,   \label{eq: PID sin filtro}
\end{equation}
where $K_p$, $K_i$, and $K_d$ are the proportional, integral, and derivative gains, respectively, and must be correctly chosen to obtain an appropriate dynamic for the closed-loop system.

Although equation \eqref{eq: PID sin filtro} is the general form of a PID controller, it is usually implemented with several additional structures, such as setpoint weighting, output signal filtering, reference filtering, and a derivative action filter, among others \citep{sundstrom2026practical}. These help avoid problems such as noise amplification due to the proportional action, high-frequency noise in the control signal caused by the derivative term, or infinite control action changes provoked by the derivative term when setpoint changes are introduced \citep{alfaro2013performance,alexis2022design,hagglund2013unified}. However, herein, the simplest form of the PID controller will be considered and analysed. 




When PID controllers are used in industry, actuator saturation must be considered, and the final control action ($u_{sat}$) is always bounded by the actuator's limits. When the controller output, $u_c$, exceeds the actuator limits, the actuator will remain at its limit, and the system runs in open loop \citep{astrom_advanced_2006}. Moreover, the integral action continues to increase as long as the error is non-zero. This causes this term to become very large, an effect known as ``integral windup''. 

\begin{figure}[h!]
    \centering
    \includegraphics[width=0.9\linewidth]{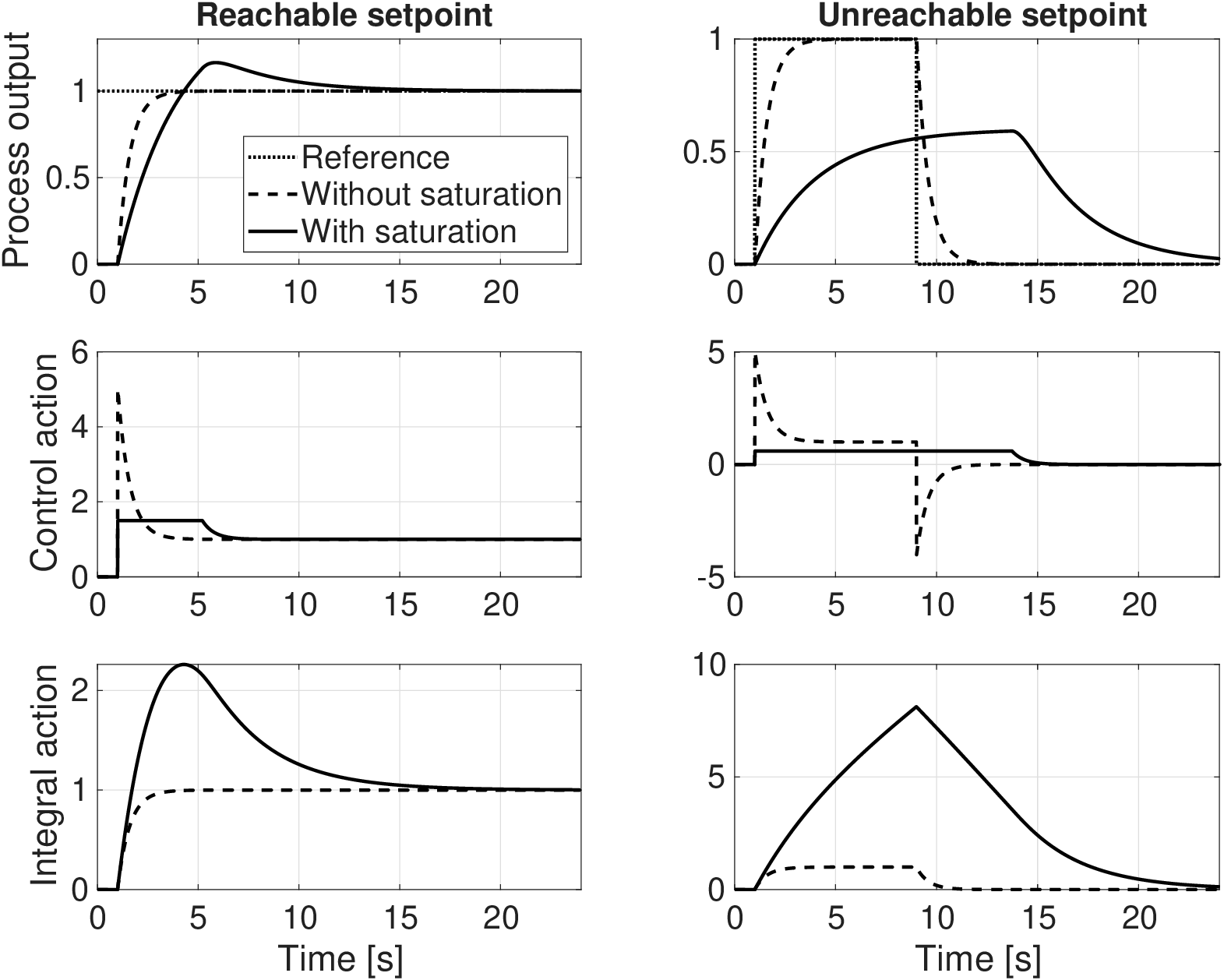}
    \caption{Representative control scenarios illustrating the effect of actuator saturation in the setpoint tracking problem. The first row shows the reference, and the process output with (solid line) and without (dashed line) saturation. The second row shows the saturated and unsaturated control actions, whilst the last row shows the integral action of both situations.}
    \label{fig: Windup effect setpoint tracking}
\end{figure}
\begin{figure}[h!]
    \centering
    \includegraphics[width=0.5\linewidth]{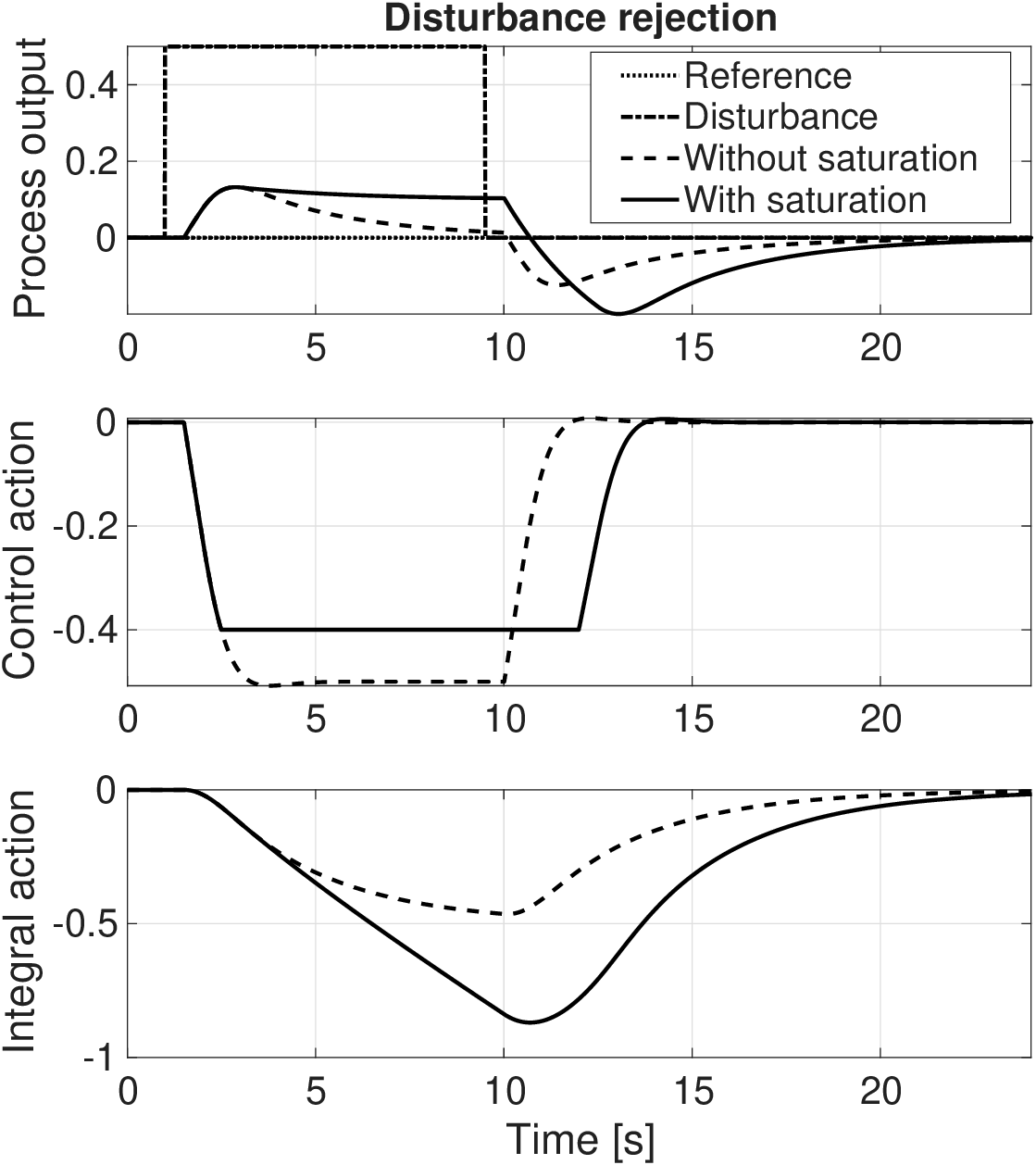}
    \caption{Representative control scenario illustrating the effect of actuator saturation in the disturbance rejection problem. The first row shows the reference and disturbance, and the process output with (solid line) and without (dashed line) saturation. The second row shows the saturated and unsaturated control actions, whilst the last row shows the integral action of both situations.}
    \label{fig: Windup effect disturbance rejection}
\end{figure}

To illustrate the impact of actuator saturation under different operating conditions, Figures \ref{fig: Windup effect setpoint tracking} and \ref{fig: Windup effect disturbance rejection} present three representative control scenarios that will be analysed throughout this work. These scenarios reflect common situations encountered in industrial practice and highlight the diverse manifestations of windup.

In all cases, the analysed process is a first-order-plus-dead-time (FOPDT) system with unitary gain $K$, three-second time constant $T$ and 0.5-second delay $L$:
\begin{equation}
    P(s) = \frac{K}{Ts+1} e^{-Ls} = \frac{1}{3s+1} e^{-0.5s}.
\end{equation}

For first-order systems, PI controllers are recommended in most cases \citep{astrom_advanced_2006,skogestad2003simple}. In the presented examples, $C(s)$ is tuned using the $\lambda$ method \citep{astrom_advanced_2006}, which, for FOPDT systems, establishes:
\begin{equation}
    \begin{array}{ll}
     K_p = \dfrac{T}{K(\lambda + L)},\\[10pt]
     K_i = \dfrac{K_p}{T_i}, \quad \text{where $T_i = T$}.
    \end{array}
\end{equation}
The desired closed-loop time-constant is $\lambda=0.2T$. 

The first two cases correspond to setpoint-tracking problems. In the example to the left in Figure \ref{fig: Windup effect setpoint tracking},  saturation occurs during the transient following a setpoint change, where the control action temporarily reaches the actuator limits but eventually returns to the linear region. This causes the control signal to grow during saturation, resulting in a slow recovery and an overshoot on the process output. In contrast, the right example in Figure \ref{fig: Windup effect setpoint tracking} shows a situation in which the actuator saturates at steady state, meaning that the required control effort to reach the reference exceeds the actuator limits, rendering the setpoint unattainable. Apart from the first setpoint becoming unreachable, when a second step is introduced, the integral action has become so high that a slow tracking is obtained and the process output does not reach the setpoint until after a long period of time.

The last case corresponds to the disturbance-rejection problem. Figure \ref{fig: Windup effect disturbance rejection} illustrates the response to a load disturbance, where the controller must counteract the disturbance, often leading to sustained saturation. The control action exceeds the actuator's limits during the disturbance period, and once this is over, the integral action has become so large that the process output recovers much more slowly than when no saturation occurs. 

Across all cases, saturation leads to a mismatch between the controller output and the applied control signal, causing the integral term to accumulate in a manner that is inconsistent with the actual system dynamics. The resulting windup effect manifests differently depending on the control objective and system configuration, motivating the need for tailored anti-windup strategies and tuning rules for each scenario.

It is clear that actuator saturation degrades system performance, and anti-windup techniques must be applied to mitigate its effects. However, selecting and tuning the anti-windup scheme is not trivial. It depends on several factors, including process and disturbance dynamics, and the controller tuning method \citep{Bohn1995Analysis}. For this reason, the following section presents a bibliographic study of several existing anti-windup techniques and discusses their advantages and disadvantages. 

\section{Anti-windup techniques review} \label{sec: review}

\begin{figure}[h!]
    \centering
    \includegraphics[width=0.7\linewidth]{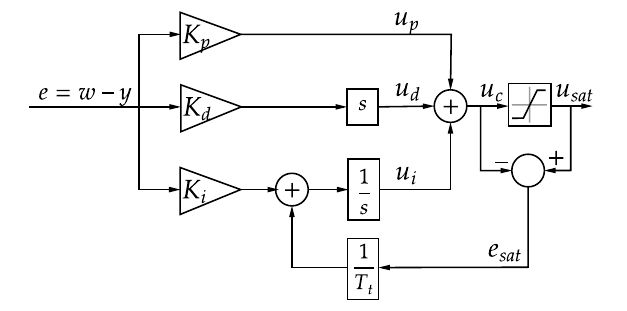}
    \caption{Classic back-calculation scheme for PID controller}
    \label{fig: Back-calculation scheme}
\end{figure}

The windup problem when the actuator saturates has motivated a wide range of anti-windup strategies, which differ in complexity, generality, and performance under saturation. The most widely implemented anti-windup strategy in industry is \textit{Back-Calculation}, also known as Tracking, in which a feedback correction is introduced to the integral term, based on the difference between the saturated and unsaturated control signal \citep{astrom_advanced_2006,peng1996anti,shin1998new}. Figure \ref{fig: Back-calculation scheme} shows the implementation of classic back-calculation applied to a PID controller, which works as follows: when the actuator saturates, the saturation error ($e_{sat} = u_{sat}-u_c$) is subtracted from the integral term weighted by a gain $1/T_t$, where $T_t$ is the tracking time constant. This way, the integral action is reset to the actuator's limit, but not instantly upon saturation. However, this feedback loop does not interfere with the PID behaviour when the controller output is within the actuator boundaries ($e_{sat} = 0$). 

In this scheme, the integral action, $u_i$ is obtained as shown in Equation \eqref{eq: calculo ui con bc}:
\begin{equation}
    u_i(s) = \left( K_i e(s) + \frac{1}{T_t} e_{sat}(s) \right) \frac{1}{s}, \label{eq: calculo ui con bc}
\end{equation}
which, when implemented in the discrete domain, may be implemented as follows:
\begin{equation}
    u_i(k) = u_i(k-1) + K_iT_se(k) + \frac{1}{T_t} T_s e_{sat}(k),      \label{eq: acción integral con anti-windup discreto}
\end{equation}
where $T_s$ is the sampling time, and $k$ is the discrete time instant. The integral term is updated at instant $k$, with its value in the previous sample, $k-1$, and the contributions produced by the error and saturation error, both multiplied by their respective gains and the sampling time. When saturation error is zero, the integral term is calculated in the normal way.

Regarding the tracking time constant, although there is no specific rule for choosing its value, several authors suggest that it should be chosen as $T_t = \sqrt{T_iT_d}$ for PID controllers and $T_t = T_i$ when there is no derivative action \citep{astrom_advanced_2006,morohoshi2025nonlinear}, being $T_i = K_p/K_i$ and $T_d = K_d/K_p$ the integral and derivative times, respectively. Other studies suggest that it should be chosen below $T_i$ \citep{rundqwist_anti-reset_1990,markaroglu_tracking_2006, Caparroz2026TuningRule}. However, a tracking time constant that is too small resets the controller too quickly, causing severe performance deterioration \citep{rundqwist_anti-reset_1990}. This parameter plays a critical role in the trade-off between fast recovery from saturation and closed-loop stability. This demonstrates the importance of the chosen value and motivates the development of rules and guidelines for tuning the tracking time constant to ensure better performance. For this reason, several guidelines are provided herein to help the reader choose the best option for their specific case based on the available information. 

Another commonly used anti-windup scheme is \textit{control signal clamping}, in which the integral term is allowed to grow up to the saturation limit \citep{sundstrom2026practicalguidepidcontroller}. Therefore, when saturation is reached, the integrator is still fed with the error, but the saturation error, $e_{sat}$, is removed from it entirely. If the scheme in Figure \ref{fig: Back-calculation scheme} is analysed, control signal clamping may also be seen as back-calculation when the saturation error feedback gain is infinite, which corresponds to $T_t=0$. In discrete implementation, this corresponds to $T_t = T_s$, if the integral action is implemented as shown in equation \eqref{eq: acción integral con anti-windup discreto}. \textcolor{black}{In this case, the integral action is immediately reset to its limit value, which can correspond to the minimum or the maximum one}. This technique is characterised by a rapid exit from actuator saturation but usually results in degraded setpoint-tracking performance. However, it has the advantage of requiring no parameters to tune, is easy to implement, and has a low computational cost. 

For this reason, herein back-calculation and control signal clamping will be referred to as ``\textit{dynamic}'' and ``\textit{instantaneous}'' back-calculation, respectively. This is mainly because both strategies back-calculate the integrator input based on the saturation error. The only difference between them is that, in dynamic back-calculation, the correction is applied using the dynamic specified by $T_t$, whilst in instantaneous back-calculation, the saturation error is immediately removed from the integral action.

In \citep{glattfelder1986start,bohn_analysis_2002}, a modified dynamic back-calculation scheme is proposed, implemented for both PI and PID controllers. In them, the authors address the problem of slow system responses that result from selecting small values of $T_t$. To mitigate this effect, an additional saturation limit is applied to the proportional-derivative actions, which are then used to generate the anti-windup feedback loop. This extra limitation allows the use of lower tracking time constants without driving the integrator too far, thereby avoiding slow responses. Moreover, this method is less sensitive to changes in system parameters, making it easier to tune and reducing the risk of performance degradation due to an incorrect parameter choice.

\textit{Conditional integration}, also known as integrator clamping, is another well-known, widely implemented anti-windup strategy in which integration is switched off when a certain condition is met \citep{astrom_advanced_2006,hanus1987conditioning,kumar2012comparative}. Some examples of conditional integration include limiting the integral term to a certain value, stopping the integration when the error exceeds a selected threshold, or when the controller saturates. In \citep{visioli_modified_2003}, conditional integration is combined with dynamic back-calculation. More specifically, dynamic back-calculation is used only when the following condition is achieved: the system error has the same sign as the manipulated variable and the system output has left its previous setpoint value. This condition allows the integral term to grow whilst the process output transient has not started. This is especially useful in processes with high dead time, but when the delay is small, this technique performs almost the same as the standard dynamic back-calculation. 

In \citep{bruciapaglia1986estrategia}, the authors propose an anti-windup strategy in which the current control signal is modified to maintain the consistency between the controller's output (unsaturated) and the applied control signal (saturated). The \textit{error recalculation} is performed by treating the control loop as a system of consistent algebraic relationships. Instead of simply clipping the controller output when it reaches a physical limit, the algorithm works backwards: it takes the actual saturated control value ($u_{sat}$) and uses the controller's internal equations to determine what the input error ($e$) and internal states would have been to produce exactly that value. This recalculated error is then used to update the controller's memory for the next time step. By effectively ``resetting'' the controller to align with the actuator's physical reality, the method ensures that the control loop remains numerically well-conditioned and ready to resume normal operation as soon as the system exits the saturation zone \citep{silva2017controle}. However, this scheme does not work for disturbance rejection and may be sensitive to modelling errors. 

Whilst classical anti-windup techniques, such as the ones mentioned above, are typically motivated by the integrator windup phenomenon, in practice, performance degradation under actuator saturation is not exclusively caused by the integral term. In particular, when additional dynamic elements are present in the control structure, other internal states may also become inconsistent with the saturated control signal, leading to undesirable transients. This may be the case for PID controllers with a derivative filter or for schemes that include feedforward or decoupling structures. 

In this sense, \citep{goodwin2001control} presents a general scheme for implementing back-calculation. In this approach, the controller is decomposed into a direct feedthrough term, $c_{\infty}$, and a strictly proper transfer function, $\overline{C}(s)$, such that the complete controller can be written as:
\begin{equation}
C(s) = c_{\infty} + \overline{C}(s),
\end{equation}
where $c_{\infty} = \lim_{s \to \infty} C(s)$ represents the high-frequency gain of the controller, and $\overline{C}(s)$ contains all its dynamic components.

Based on this decomposition, the anti-windup mechanism is constructed by feeding back the difference between the applied control signal through a dynamic compensator defined by:
\begin{equation}
[C(s)]^{-1} - c_{\infty}^{-1}.
\end{equation}

\begin{figure}[h!]
    \centering
    \includegraphics[width=0.6\linewidth]{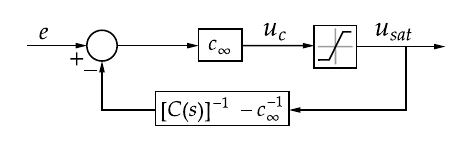}
    \caption{Anti-windup scheme presented in \citep{goodwin2001control}}
    \label{fig: Esquema Goodwin}
\end{figure}

This results in a generalised back-calculation scheme in which the correction term acts on the full controller dynamics rather than only on the integral component \citep{kothare1994unified,hanus1987conditioning,walgama1992generalisation,walgama1990inherent}. The resulting structure, illustrated in Figure \ref{fig: Esquema Goodwin}, allows the control action to be constrained not only in amplitude (saturation) but also in rate (slew-rate limits), whilst maintaining consistency with the internal dynamics of the controller.

One of the main advantages of this formulation is its generality. Unlike classical back-calculation, which implicitly assumes that the integrator is the sole source of windup, this approach applies to controllers of arbitrary order, including PID controllers with filtered derivative action and more complex compensators. As a result, all internal states of the controller are properly driven by the saturated control signal, preventing the accumulation of hidden dynamics that may lead to undesirable transients when the actuator exits saturation. Additionally, this method does not require the tuning of a tracking parameter, as the anti-windup compensator is directly derived from the controller itself. This provides a systematic and theoretically consistent design procedure.

However, these benefits come at the cost of increased implementation complexity. The need to compute and realise the controller's inverse dynamics may introduce numerical and practical difficulties, particularly in discrete-time implementations or when the controller model is high-order or uncertain. Moreover, the lack of a tuning parameter, whilst theoretically appealing, reduces the flexibility to shape the transient response during saturation, a feature that can be advantageous in practical applications when using classical back-calculation schemes.

In the specific case of feedforward schemes, the classic anti-windup limitation is addressed in \citep{hoyo_practical_2023}, which proposes an anti-windup scheme for control architectures that include feedforward compensation. This approach extends the anti-windup mechanism beyond the feedback controller, ensuring consistency between the saturated actuator signal and all contributing control paths. In this framework, both the feedback and feedforward components are taken into account when designing the anti-windup compensator, preventing the mismatch that arises when the feedforward action continues to drive the actuator as if no saturation were present. As a result, the proposed method avoids the accumulation of hidden dynamics not only in the integrator but also in other parts of the control structure.

Overall, the reviewed techniques can be interpreted as progressively enforcing consistency at different levels of the control system. Classical methods focus on the integrator state; generalised back-calculation extends this correction to the full controller dynamics; and more advanced approaches incorporate multiple control paths, such as feedforward actions. This progression reflects a trade-off between implementation simplicity and the ability to handle increasingly complex control architectures under saturation. However, this work will focus on anti-windup techniques for the classical PI controller, in which the windup effect is caused exclusively by the integrator. In the following section, several previously mentioned anti-windup techniques will be described in depth, with the ultimate objective of comparing them across different situations and deriving conclusions and guidelines to help the users choose the best anti-windup solution for their specific problem.




\section{Compared anti-windup schemes} \label{sec: compared solutions}
Following the review presented in the previous section, this section focuses on a subset of anti-windup strategies that will be analysed and compared in detail through simulation studies. The selection of these methods is guided not only by their theoretical relevance but also by their widespread adoption in industrial practice. 

The schemes described in the following sections will be compared for FOPDT systems with PI controllers tuned using the $\lambda$ method and evaluated under a wide variety of operating conditions.


The comparisons will include setpoint-tracking and disturbance-rejection problems, as shown in Section \ref{sec: windup}. Moreover, examples under \textcolor{black}{different saturation conditions} and controller tunings will be presented. The final aim is to determine which anti-windup technique is preferable to apply depending on the system dynamics, the controller, and the control problem. 

The following subsections describe each of these techniques in detail, providing the necessary formulations and implementation details for a fair and consistent comparison.

\subsection{Classic schemes}    \label{subsec: classic schemes}

In particular, classical anti-windup techniques such as dynamic back-calculation and instantaneous back-calculation remain the most commonly implemented solutions in industrial PID controllers, due to their simplicity, low computational cost, and ease of integration within existing control architectures. These anti-windup schemes were described above and presented in Figure \ref{fig: Back-calculation scheme}. As mentioned in Section \ref{sec: review}, both anti-windup schemes are analogous, but, in dynamic back-calculation, the value of $T_t$ must be chosen by the user and, in instantaneous back-calculation, the tracking time constant must be equal to zero, in the continuous domain, and equal to the sampling time, $T_s$, in the discrete domain. 

For the dynamic back-calculation scheme, several $T_t$ tuning rules will be tested. The first is the well-known rule of thumb in which $T_t = T_i$, suggested by several authors \citep{astrom_advanced_2006}. This can be implemented as shown in Algorithm \ref{alg: PI anti-windup BC}, using the same code as in instantaneous back-calculation, by just changing the value of $T_t$. 

\begin{algorithm}[!ht]
\caption{PI Controller with back-calculation anti-windup}
\label{alg: PI anti-windup BC}
\begin{algorithmic}[1]
\State \textbf{Initialization:}
\State Choose sampling time: $T_s$
\State Choose controller tuning: $K_p$, $K_i$
\State Choose anti-windup tuning: $T_t$
\State \textbf{for} each sampling instant $k$ \textbf{do}
    \State \Indent \textbf{1. Measurement}
    \State \Indent Define reference $w(k)$
    \State \Indent Measure output $y(k)$
    \State \Indent Compute error $e(k)$
    
    \State  \Indent \textbf{2. Control action computation}
    \State  \Indent Compute proportional action $u_p(k) = K_p e(k)$
    \State  \Indent Compute integral action $u_i(k)$ with back-calculation anti-windup:
    \[
    u_i(k) = u_i(k-1) + K_i T_s e(k)  + \frac{1}{T_t} T_s e_{sat}(k-1) 
    \]
    \State  \Indent Compute total control action $ u_c(k) = u_p(k)+ u_i(k)$
    \State  \Indent Apply actuator saturation $u_{sat}(k) = \min ( \max (u_c(k),u_{min}),u_{max})$
    \State  \Indent Compute saturation error $ e_{sat}(k) = u_{sat}(k) - u_c(k)$
    \State  \Indent Apply control action $u_{sat}(k)$
\State \textbf{end for}
\end{algorithmic}
\end{algorithm}

\begin{figure}[h!]
    \centering
    \includegraphics[width=0.7\linewidth]{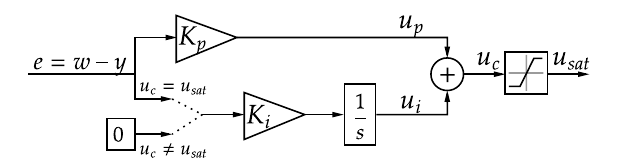}
    \caption{Conditional integration anti-windup scheme}
    \label{fig: Esquema Conditional Integration}
\end{figure}

Conditional integration strategies are also frequently employed, especially in applications where straightforward logic-based solutions are preferred. In this case, the corresponding scheme is shown in Figure \ref{fig: Esquema Conditional Integration}, where a switch governs the input to the integrator based on the saturation error. 

To implement this scheme, the integral action must be computed as shown in Algorithm \ref{alg: Conditional integration}, which is the only change required to Algorithm \ref{alg: PI anti-windup BC} in line 12. Although this scheme could be implemented with an if/else statement (option 1), a one-line alternative is also proposed to enable faster code evaluation (option 2). Note that, in the second option, the expression between bigger parentheses returns a boolean value, allowing integration as long as saturation error is zero, meaning $u_c = u_{sat}$. When this condition is not met, the result of this evaluation is zero, and the integral term remains with the previous value.  

\begin{algorithm}[!t]
\caption{Modification to integral action computation for conditional integration anti-windup}
\label{alg: Conditional integration}
\begin{algorithmic}[1]
\State Compute integral action $u_i(k)$ with conditional integration:
\begin{center}
\begin{minipage}[t]{0.45\linewidth}
    {\small  
    \begin{algorithmic}[0]
    \State \textit{Option 1:}
    \IF{$e_{sat}(k-1) == 0$}
        \STATE $u_i(k) = u_i(k-1) + K_i T_s e(k)$
    \ELSE
        \STATE $u_i(k) = u_i(k-1)$
    \ENDIF
    \end{algorithmic}
    }
\end{minipage}
\hfill\vrule\hfill
\begin{minipage}[t]{0.45\linewidth}
    {\small  
    \begin{algorithmic}[0]
    \State \textit{Option 2: }
    \begin{align*}
        u_i(k) & = u_i(k-1) + ...\\
        & \bigg(e_{sat}(k-1) == 0\bigg)  K_i  T_s e(k)
    \end{align*}
    \end{algorithmic}
    }
\end{minipage}
\end{center}
\end{algorithmic}
\end{algorithm}

\subsection{Hybrid schemes}

In addition to the standard approaches, hybrid schemes that combine the advantages of multiple techniques have been proposed to improve performance under saturation. Among these, the method introduced in \citep{visioli_modified_2003}, which integrates conditional integration with back-calculation, represents a practical compromise between simplicity and effectiveness. Therefore, the corresponding scheme is a combination of the ones shown in Figures \ref{fig: Back-calculation scheme} and \ref{fig: Esquema Conditional Integration}, in which the tracking time constant suggested by the authors is $T_t=0.03T_i$ and the integral action is obtained as follows:
\begin{equation}
    u_{i}(k) = 
    \begin{cases} 
    u_i(k-1) + \dfrac{K_{p}}{T_{i}} T_se(k) + \dfrac{1}{T_{t}} T_se_{sat}(k-1), & \text{if }condition\text{,}\\[10pt] 
    u_i(k-1) + \dfrac{K_{{p}}}{T_{{i}}} T_s e(k), & \text{otherwise,} 
    \end{cases}
\end{equation}

\noindent where \textit{condition} is satisfied when the control signal has saturated, the system error has the same sign as the manipulated variable, and the system output has left its previous setpoint value. Mathematically, this can be expressed as follows:
\begin{align}
    \left[u(k-1) \neq u_{sat}(k-1)\right] \wedge \left[u(k-1) \times e(k-1) > 0\right]  \\
    \wedge
    \begin{cases} 
    y(k) > y(k-2) & \text{if } y(k-1) > y(k-2) \\ 
    y(k) < y(k-2) & \text{if } y(k-1) < y(k-2).
    \end{cases}
\end{align}

\subsection{Other tuning rules for back-calculation}    \label{subsec: other solutions}


As discussed in Section \ref{sec: review}, in addition to classical solutions, several authors have proposed new strategies to enhance controller performance under actuator saturation.  

The dynamic back-calculation scheme's performance will be compared with other tuning rules, including the well-known rule of thumb. In the case of setpoint tracking problems, the $T_t$ tuning rule proposed in \citep{markaroglu_tracking_2006} will be tested. This rule exploits the advantage of choosing big and small values for this constant. When the controller output saturates, the value of $T_t$ is chosen big enough to allow a very fast increase in the process output ($T_t^1 = 10 T_i$), while remaining in saturation for longer. When the measured output reaches a certain percentage value $c$ of the system reference, $T_t$ is decreased to a value $T_t^{new} = \beta T_i$, avoiding big overshoots as the integral term is suppressed. The value of $c$ depends on the open-loop static gain, $K$, and the maximum value of the saturated output, $u_{max}$, and it can be obtained as follows:
$$
c = \begin{cases} 
\left(-0.5\,\dfrac{u_{\max} K}{w} + 1.4\right)100 & \text{for} \quad 1 \leq R_c \leq 2.6, \\
10 & \text{for} \quad 2.6 < R_c ,
\end{cases}
$$
where $R_c = u_{\max}K/w$. Moreover, $\beta$ depends on the $T/L$ ratio, and is obtained using the following equation:
\begin{equation}
    \beta = \left(0.59 - 0.65 \cdot e^{-0.09 \frac{T}{L}} \right). \label{eq: Tt new}
\end{equation}  

Therefore, for this anti-windup implementation, some modification is required to Algorithm \ref{alg: PI anti-windup BC} within the control loop: after measuring, before computing the control action, $T_t$ must be computed. This means the error percentage $c$ must be checked in every iteration, and $T_t$ must be selected. However, if the reference changes are known a priori, $R_c$ and $c$ can be computed outside the loop, being only necessary to check the difference between the output and the reference in order to select between both values of $T_t$, which can also be computed outside the loop. Although the computation of $c$ and the switch between $T_t^1$ and $T_t^{new}$ can be implemented with if/else statements, a more compact way is shown in Algorithm \ref{alg: BC changing Tt}. As in Algorithm \ref{alg: Conditional integration}, when computing $c$ and $T_t$, the expressions between bigger parentheses consist of boolean expressions that will provide 1 or 0 as results. As mentioned before, this may be added to Algorithm \ref{alg: PI anti-windup BC} inside or outside the control loop, depending on the available information regarding the reference. 

\begin{algorithm}[h!]
\caption{Modification to PI Controller with back-calculation anti-windup changing $T_t$}
\label{alg: BC changing Tt}
\begin{algorithmic}[1]
    \State  \Indent \textbf{Compute $c$ and choose $T_t$}      
    \[
    R_c = u_{max} K/ w(k)
    \]
    \[
    c = \bigg(R_c>2.6\bigg) 0.1 + \bigg(R_c\leq1\bigg) + \bigg(R_c >1 \text{ } \&\& \text{ }R_c <=2.6\bigg) (-0.5 R_c+1.4)
    \]
    \[
    T_t = \bigg(y(k) \leq w(k) c\bigg) T_t^1 + \bigg(y(k) > w(k)  c\bigg) T_t^{new}
    \]
\end{algorithmic}
\end{algorithm}

On the other hand, for disturbance rejection problems, new tuning rules for $T_t$ are proposed and described in depth in section \ref{sec: New solutions}, for load disturbances in schemes without a feedforward compensator.

\section{New proposed solutions}    \label{sec: New solutions}


In this section, new ideas for addressing integrator windup are proposed. Whilst the first solution suggests a new anti-windup strategy that combines existing ones, the second provides new tuning rules for the tracking time constant used in back-calculation schemes, based on an optimisation procedure.

\subsection{New hybrid anti-windup scheme}    \label{subsec: Kristians idea}
In this section, a new method is proposed, in which two complementary anti-windup mechanisms are applied sequentially within each control cycle. After computing the unsaturated controller output, the algorithm first checks whether the saturation error, $e_{sat}$, has the opposite sign to the current integral increment, $\Delta u_i = K_iT_se(k)$. If this condition holds, it indicates that the integrator is actively driving the control signal further into saturation. If that is the case, a direct correction is applied: the controller output is shifted by the smaller of the absolute saturation excess and the absolute integral step, in the direction that reduces the excess. This conditional clipping effectively prevents the integrator from accumulating beyond the saturation boundary in real time, without interfering with integral action when the system operates within its linear range.

Following this conditional correction, a classical dynamic back-calculation term is applied, adding the saturation error multiplied by the corresponding gain, $1/T_t$, to the integral term. This provides a tunable, smooth discharge of any remaining windup via the tracking time constant. The combination results in a two-stage correction: the first stage is aggressive and instantaneous, reacting within the same sample to large saturation events, whilst the second stage governs recovery dynamics in a smooth and adjustable manner. Together, they aim to keep the integrator state close to the saturation boundary rather than allowing it to drift far beyond it.

The main advantage of this hybrid approach is its responsiveness: the conditional clipping stage reacts immediately to windup onset without requiring tuning, whilst the back-calculation stage offers the designer a degree of freedom to shape recovery behaviour, making the method more robust to sudden large setpoint changes than back-calculation alone. 

It must be noted that, to implement this anti-windup scheme, an incremental PI algorithm is required. Therefore, Algorithm \ref{alg: PI anti-windup hybrid} shows the control action computation. Note that, when applying the second stage of the anti-windup scheme, one may choose between dynamic back-calculation, by choosing a value of $T_t>T_s$, or instantaneous back-calculation, by choosing $T_t<T_s$.

\begin{algorithm}[!ht]
\caption{Incremental PI controller with hybrid anti-windup}
\label{alg: PI anti-windup hybrid}
\begin{algorithmic}[]
\State \textbf{for} each sampling instant $k$ \textbf{do}
    \State \Indent \textbf{1. Measurement}    
    \State  \Indent \textbf{2. Control action computation}
    \State  \Indent Compute proportional action increment $du_p = K_p (e(k)-e(k-1))$
    \State  \Indent Compute integral action increment $du_i = K_i T_s e(k)$
    \State  \Indent Compute total control action $ u_c(k) = u_c(k-1) + du_p+ du_i$
    \State  \Indent Apply actuator saturation $u_{sat}(k) = \min ( \max (u_c(k),u_{min}),u_{max})$
    \State  \Indent Compute saturation error $ e_{sat}(k) = u_{sat}(k) - u_c(k)$
    \State \Indent Check integral action direction and correct:
    \begin{algorithmic}
        \IF{$\text{sign}(e_{sat}(k)) ==\text{sign}(du_i(k))$}
            \STATE $u_c(k) = u_c(k) - \text{sign}(e_{sat}(k)) \min(\text{abs}(e_{sat}(k)),\text{abs}(du_i(k)))$
        \ENDIF
    \end{algorithmic}
    \State Apply anti-windup's second stage: $u_c(k) = u_c(k) - \min(T_s/T_t,1) e_{sat}(k)$
    \State Apply actuator saturation $u_{sat}(k) = \min ( \max (u_c(k),u_{min}),u_{max})$
    \State  Apply control action $u_{sat}(k)$
\State \textbf{end for}
\end{algorithmic}
\end{algorithm}

\subsection{Tuning rules for the tracking time constant in back-calculation for load-disturbance rejection}      \label{subsec: New Tt rules}

\begin{figure}[h!]
    \centering
    \includegraphics[width=0.7\linewidth]{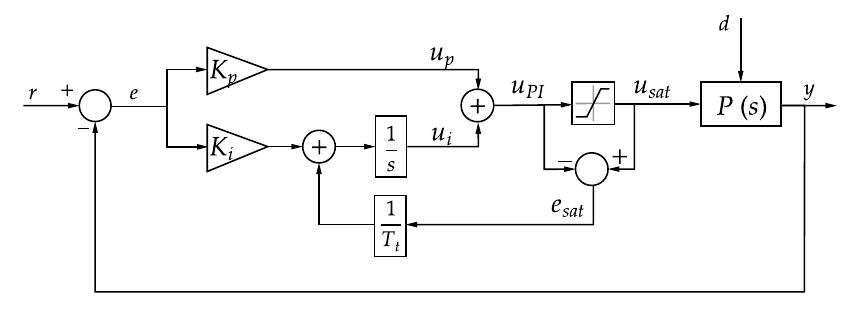}
    \caption{PI with back-calculation anti-windup scheme for $T_t$ tuning rules development}
    \label{fig: Scheme Tt rules general}
\end{figure}

Finally, this section includes a set of tuning rules for the tracking time constant in back-calculation schemes. Despite the extensive use of this technique in industry, the selection of the tracking parameter is often based on heuristic guidelines. The proposed rules aim to provide a more systematic, performance-oriented tuning procedure, particularly in scenarios where disturbance rejection is the primary objective. Therefore, the analysed scheme is the one shown in Figure \ref{fig: Scheme Tt rules general}, where the study is performed for $w=0$ and considering $d$ like a double-step disturbance signal. 

\begin{figure}[h!]
    \centering
    \includegraphics[width=\textwidth]{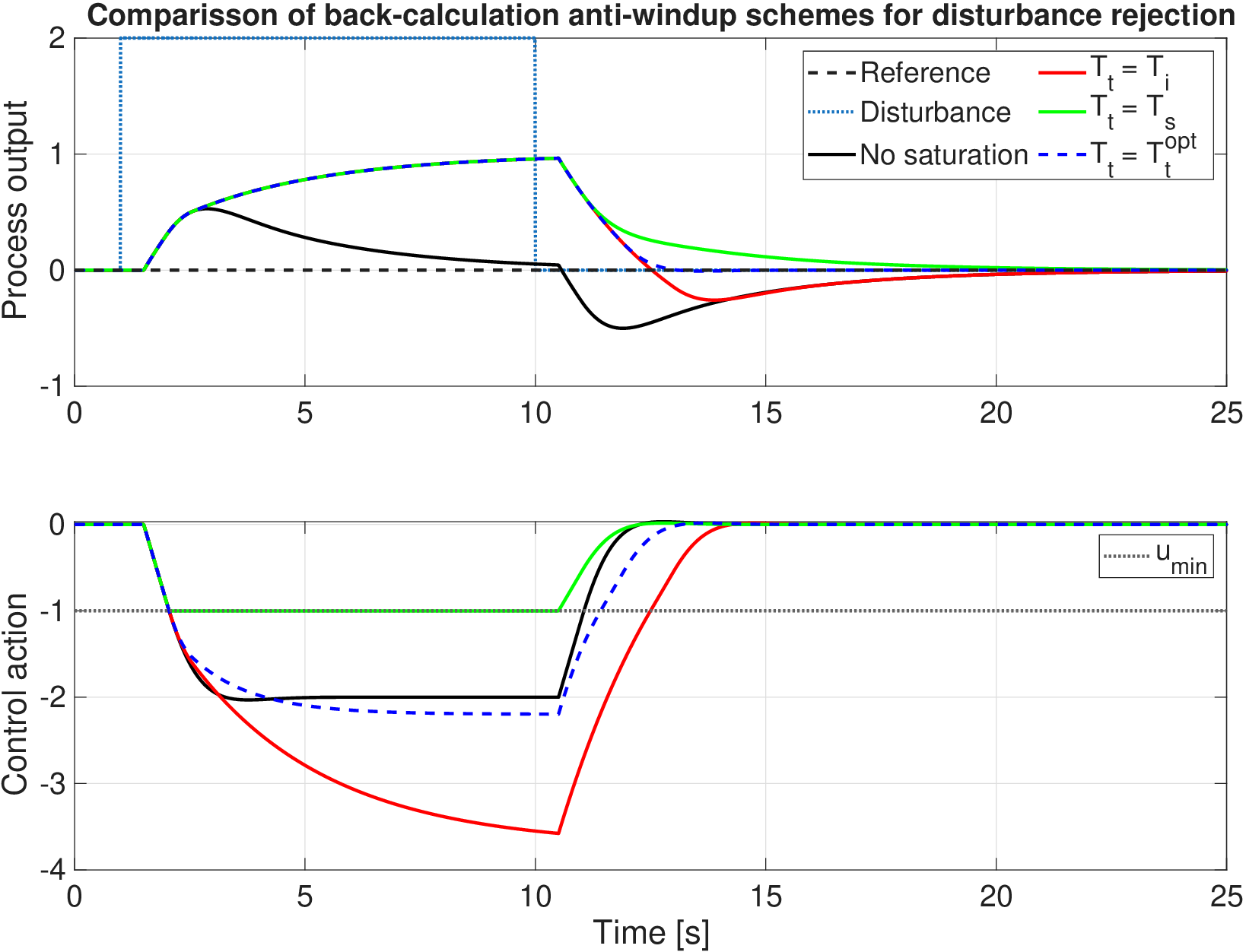}
    \caption{Back-calculation anti-windup comparison for disturbance rejection using different values of $T_t$. The first plot shows the reference (in dashed black), the disturbance (in dotted light-blue), the process output without saturation (solid black), and the process output with saturation and back calculation for three different values of $T_t$ ($T_i$ in solid red, $T_s$ in solid green, and its optimal value, $T_t^{opt}$ in dashed blue). The second plot shows the calculated control action for each of the mentioned strategies before the saturation, and the actuator's inferior limit, $u_{min} = -1$, in dotted black.}
    \label{fig: anti-windup comparison - motivation}
\end{figure}

In Figure \ref{fig: anti-windup comparison - motivation}, an example is shown to motivate the improvement achieved by the proposed tuning rule. The system performance is compared for cases with no saturation, with saturation and dynamic back-calculation ($T_t=T_i$), with saturation and instantaneous back-calculation ($T_t = T_s$), and with saturation and dynamic back-calculation using an optimal tracking time constant ($T_t = T_t^{opt}$). As mentioned previously, the disturbance is double-step-shaped, and as long as it persists, the control action remains saturated. Therefore, the developed rules are expected to enable the system to return faster to the operating point once the disturbance has subsided.

Several rules and guidelines are proposed for the user to apply, depending on the available information. The first step is to characterise the system and gather information about it. The variables that will allow the user to choose a better tracking time constant are the saturation ratio, $R_S$, the controller aggressiveness, $x$, and the disturbance-to-process-dynamics ratio. The saturation ratio is a quantitative measurement of how much the actuator is saturating, and is defined as follows:
\begin{equation}
    R_S = \frac{u_f-u_{lim}}{u_f-u_{0}},
\end{equation}
where $u_f$ is the final value of the control action without saturation when the disturbance is applied, $u_{lim}$ is the saturation limit, and $u_{0}$ is the operating point before the disturbance \citep{hoyo_practical_2023}. 

The controller's aggressiveness is the ratio of the closed- to open-loop time constants, $x = \lambda/T$, where $\lambda$ is the closed-loop time constant. Finally, the disturbance-to-process-dynamics ratio is the relationship between the pulse duration and the open-loop time constant, $D_d/T$, where $D_d$ is the pulse duration. As this information may not always be available, several versions of the rules will be offered to the reader to choose from. 

The objective of the rules is to find an expression for a factor $\alpha$, so that the tracking time constant may be computed as $T_t = \alpha T_i$, which allows a better performance. With this in mind, an optimisation problem was launched for a wide range of situations, in which the characteristic parameters were varied. The study includes values of $R_S$ from 0.05 to 0.95, $x$ from 0.2 to 1, and $D_d/T$ from 1/3 to 10. This procedure created several surfaces and a three-dimensional space to which a curve could be fitted, depending on all the parameters. Finally, this expression was simplified to yield a simple solution when not all the information is available.



\begin{figure}[h!]
    \centering
    \includegraphics[width=\linewidth]{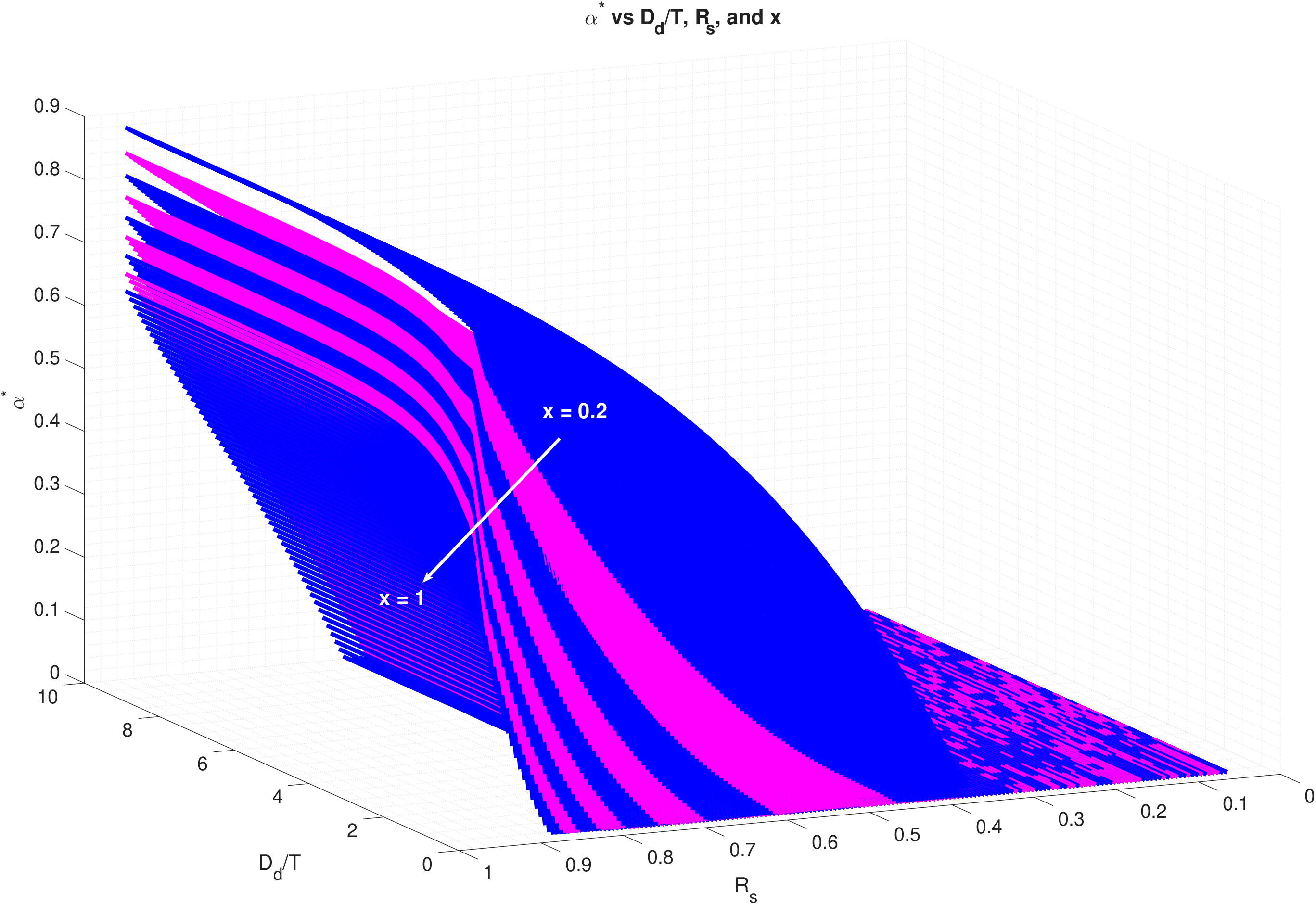}
    \caption{Optimal surfaces of $\alpha$ for load disturbance scenario}
    \label{fig: surface alpha load disturbance}
\end{figure}

The resulting surface for the optimal value of $\alpha$ ($\alpha^*$)  is shown in Figure \ref{fig: surface alpha load disturbance}, where each surface represents a value of $x$. In general, $\alpha^*$ is always less than 1, meaning the tracking time constant should always be smaller than the integral time. Moreover, for small values of $R_S$, $\alpha^* = T_s/T_i$, meaning that $T_t = T_s$ and the optimal strategy is instantaneous back-calculation. For medium to high values of $R_S$, the optimal values of $\alpha$ vary between $T_s/T_i$ and 0.9, approximately. Lastly, as the controller's aggressiveness decreases (higher $x$), the values of $\alpha$ should be smaller.  

Although several rules can be obtained from these surfaces, depending on the information available, $\alpha$ can always be computed as
\begin{equation}
    \alpha = \max\left\{f(R_S,D_d,T,x),  \frac{T_s}{T_i}       \right\},
\end{equation}
where $f$ varies from one rule to the other. A preliminary version of these rules was presented in \citep{Caparroz2026TuningRule}, where four rules were derived. Herein, for the sake of readability, only two will be compared: the most complete and the simplest, along with a simple guideline.

\textbf{Rule 1.} The first rule makes use of all the parameters that characterise the system, and $f$ is obtained as follows:
\begin{equation}
    f(R_S,D_d,T,d_x) = -1.2 + 3.3 (R_S - d_x) - 1.26 (R_S - d_x)^2 - 0.6 e^{-1.2 D_d /T}, \label{ec: Rule 1}
\end{equation}
where $d_x(x) = -0.28 + 0.8 x - 0.3 x^2$ represents the displacement on the $R_S$ axes that the surfaces experience for different values of $x$. 

\textbf{Rule 2.} The second rule analysed in this work is useful for cases in which no prior information is known about the disturbance duration, and, in this case, 
\begin{equation}
    f(R_S,x) = -0.3 -0.63 x + 1.5 R_S. \label{ec: Rule 2}
\end{equation}

\begin{figure}[h!]
    \centering
    \begin{subfigure}[b]{0.5\textwidth}
        \centering
        \includegraphics[width=\textwidth]{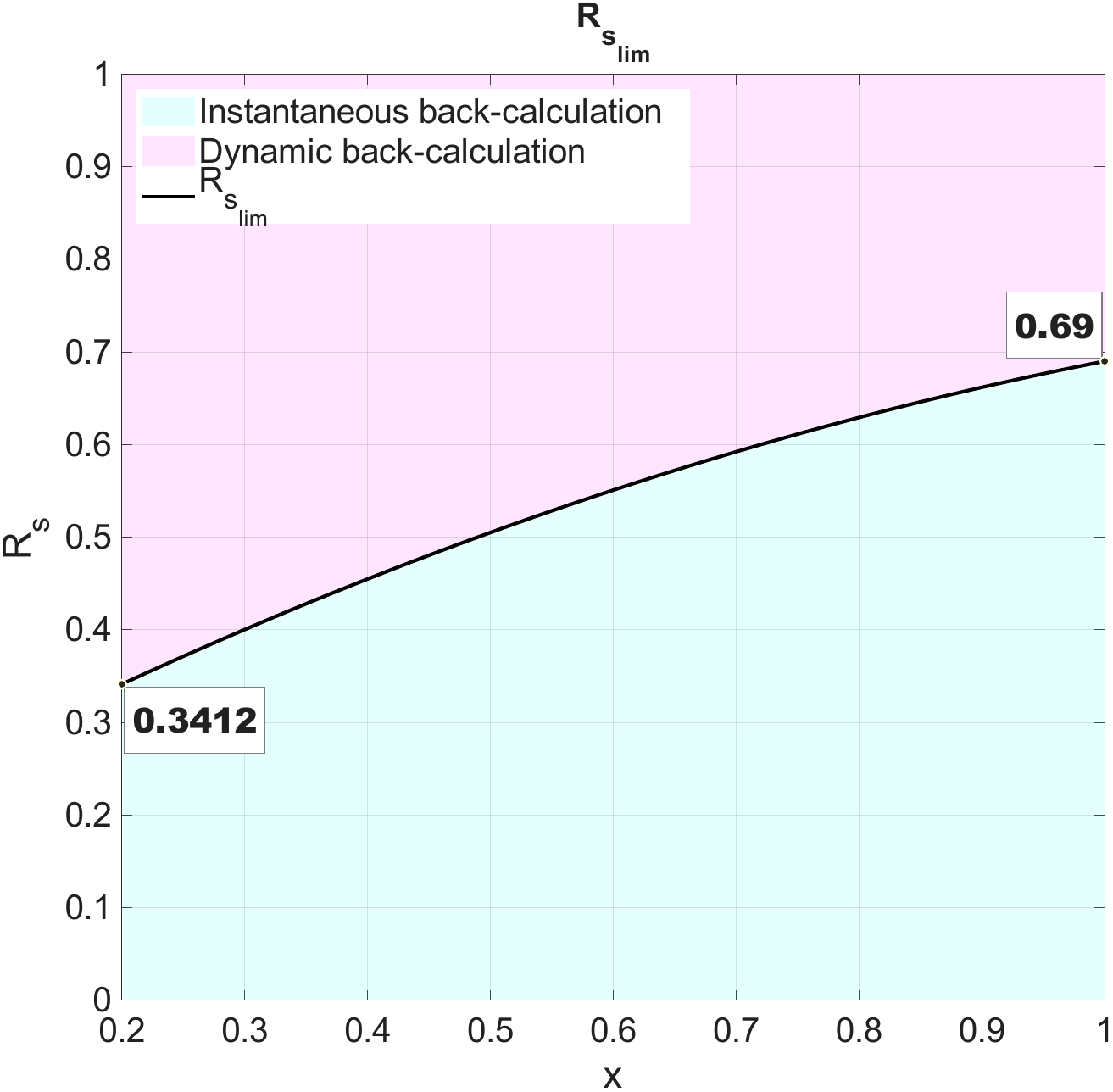}
        \caption{}
        \label{fig: Rs x 1}
    \end{subfigure}
    \hfill
    \begin{subfigure}[b]{0.49\textwidth}
        \centering
        \includegraphics[width=\textwidth]{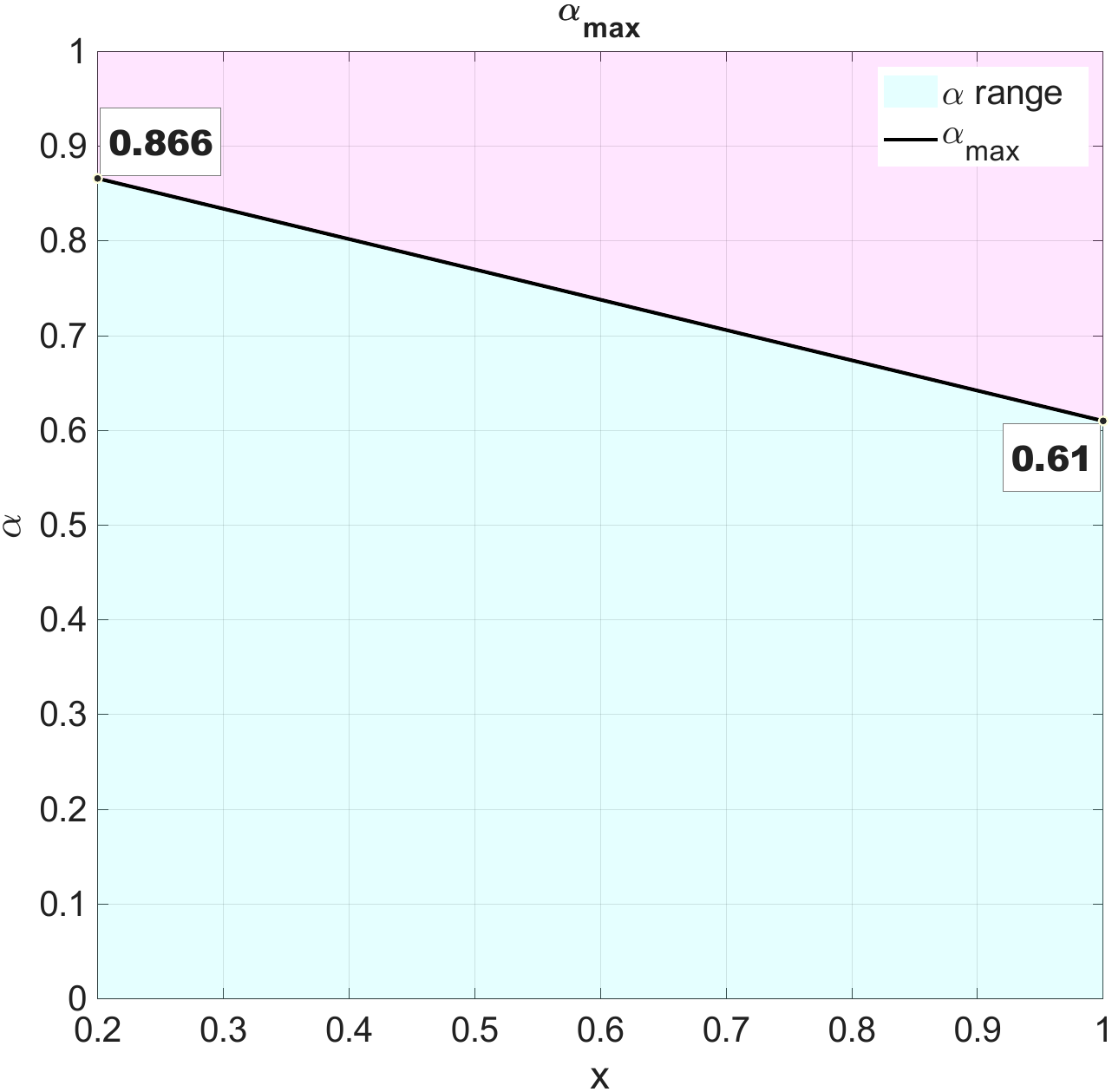}
        \caption{}
        \label{fig: alpha x 1}
    \end{subfigure}
    \caption{Limits of $R_S$ and $\alpha_{max}$}
    \label{fig: simple guideline x}
\end{figure}

\textbf{Simple guideline.} Lastly, a simple guideline is proposed for cases where only an estimate of $R_S$ can be made, in which case the previous rules cannot be applied. In Figure \ref{fig: surface alpha load disturbance}, it can be seen that each value of $x$ presents a value of $R_S$ under which $\alpha^* = T_s/T_i$. Above this value, which will be referred to as $R_{S_{lim}}$, $\alpha^*$ takes values between $T_s/T_i$ and a maximum value, $\alpha_{max}$, which is always less than the unit. Figure \ref{fig: simple guideline x} shows the curve of both limits as a function of $x$. Figure \ref{fig: Rs x 1} shows how, for saturation ratios below $R_{S_{lim}}$, the optimal anti-windup strategy is instantaneous back-calculation, as the optimisation problem results in $\alpha^* = T_s/T_i$. Above these values, on the other hand, the user can choose values of $\alpha$ below the curve of $\alpha_{max}$, in Figure \ref{fig: alpha x 1}, and, despite not obtaining the optimal behaviour, still get an important improvement compared to using $\alpha=1$.  

\section{Strategies comparison}   \label{sec: validation examples}



In this section, all the anti-windup techniques and tuning rules described in Sections \ref{sec: compared solutions} and \ref{sec: New solutions} will be compared under a wide variety of conditions, aiming to find simple guidelines that allow the user to select the best anti-windup configuration in each case.  For simplicity, Table \ref{tab: strategies codification} shows the studied strategies and the codification with which they will be referred to from now on. 

\begin{table}[h!]
\centering
\begin{tabular}{ll}
\toprule
\textbf{Strategy name}                                                                      & \textbf{Codification}     \\ 
\midrule
Dynamic back-calculation ($T_t=T_i$)                                                        & DBC1                      \\
Instantaneous back-calculation                                                              & IBC                       \\
Conditional integration                                                                     & CI                        \\
Conditional integration and back-calculation \citep{visioli_modified_2003}                   & H1                        \\
Conditional integration and back-calculation proposed                                       & H2                        \\
Dynamic back-calculation, rules by \citep{markaroglu_tracking_2006}       & DBC\_STr                  \\
Dynamic back-calculation, rule 1 for disturbance rejection                                  & DBC\_R1                   \\ 
Dynamic back-calculation, rule 2 for disturbance rejection                                  & DBC\_R2                   \\ 
\bottomrule
\end{tabular}
\caption{Strategies codification}
\label{tab: strategies codification}
\end{table}

As mentioned previously, both setpoint-tracking and disturbance-rejection problems will be studied. Moreover, setpoint tracking includes problems in which the control action is saturated during the transient and on the steady state. To obtain conclusions as general as possible for users to apply to their processes, three types of processes were analysed in which the $T/L$ ratio took the values 1/6, 1/2, and 1. Moreover, the parameters were varied, in each case, as shown in Table \ref{tab: parametros barridos}, to ensure that a wide variety of conditions are studied. 

\begin{table}[h!]
\centering
\begin{tabular}{ccll}
\toprule
\multicolumn{1}{c}{\textbf{Parameter}} & \multicolumn{3}{c}{\textbf{Values}}   \\ \midrule
\multicolumn{4}{c}{\textbf{Setpoint tracking - Saturation on the transient period}} \\ 
\multicolumn{1}{c}{\textbf{\(x\)}}         & \multicolumn{3}{c}{0.2; 0.5; 0.8}     \\ 
\multicolumn{1}{c}{\textbf{\(R_S\)}}      & \multicolumn{3}{c}{From $R_{s_{min}}$ to 0.95} \\ \midrule
\multicolumn{4}{c}{\textbf{Setpoint tracking - Saturation on the steady-state}} \\ 
\multicolumn{1}{c}{\textbf{\(x\)}}         & \multicolumn{3}{c}{0.2; 0.5; 0.8}     \\ 
\multicolumn{1}{c}{\textbf{\(R_S\)}}      & \multicolumn{3}{c}{From 0.05 to 0.95} \\ \midrule
\multicolumn{4}{c}{\textbf{Disturbance rejection}}          \\ 
\multicolumn{1}{c}{\textbf{\(x\)}}         & \multicolumn{3}{c}{0.2; 0.5; 0.8}     \\ 
\multicolumn{1}{c}{\textbf{\(R_S\)}}      & \multicolumn{3}{c}{0.35; 0.55; 0.8}     \\ 
\multicolumn{1}{c}{\textbf{\(D_d/T\)}}    & \multicolumn{3}{c}{From 1/3 to 10}    \\ \bottomrule
\end{tabular}
\caption{Studied conditions for anti-windup schemes comparison}
\label{tab: parametros barridos}
\end{table}

Lastly, all the strategies are compared in terms of output performance, for which the IAE (Integral of the Absolute Error) is computed, which is defined as follows: 
\begin{equation}
    IAE = \sum_{k=1}^{N} \left| e(k) \right|,
\end{equation}
where $k$ is the actual time instant, that goes from $1$ to the simulation length, $N$. 

\subsection{Setpoint-tracking problem}

\begin{figure}[h!]
    \centering
    \includegraphics[width=0.9\textwidth]{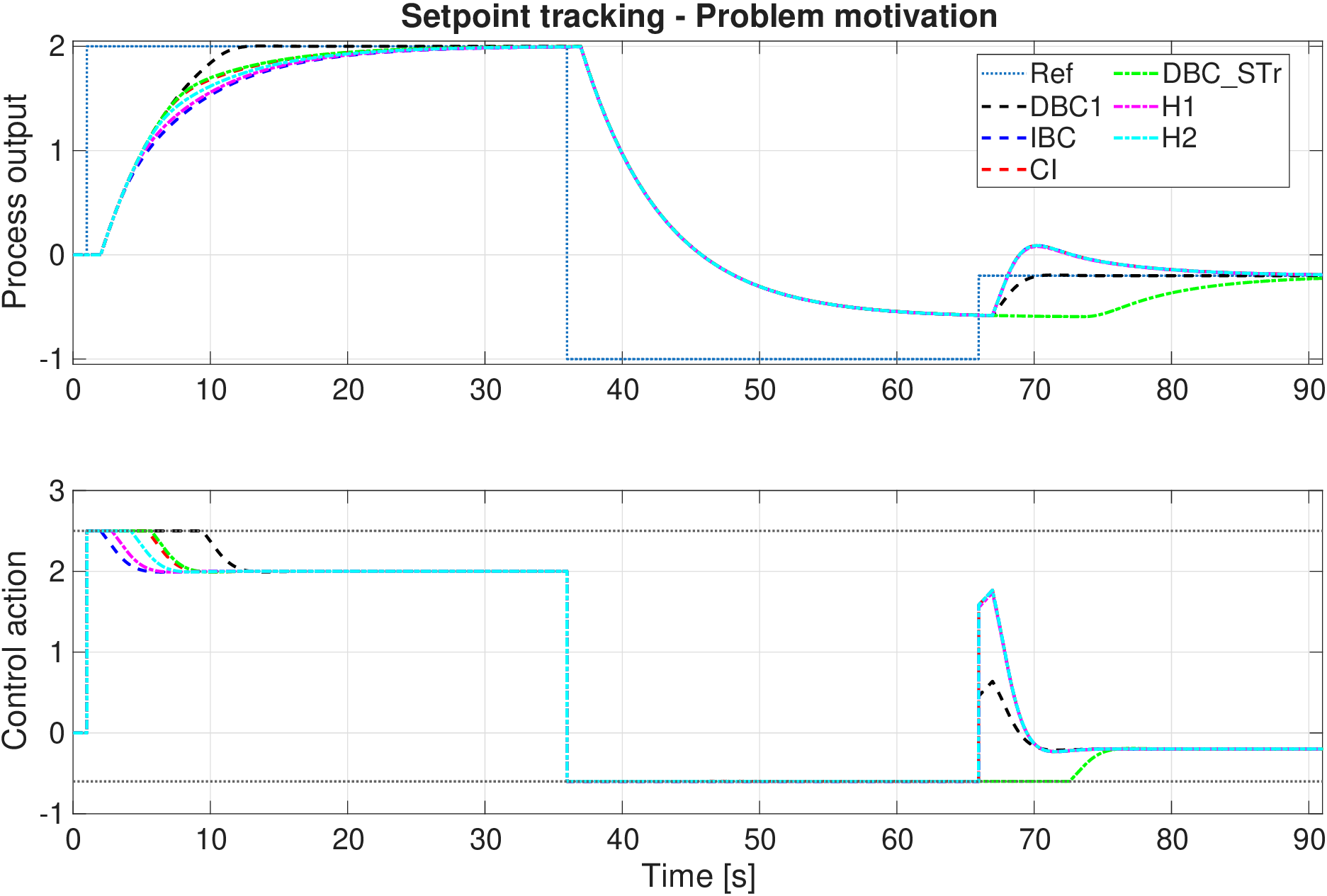}
    \caption{Analysed situations in setpoint-tracking problems}
    \label{fig: Setpoint tracking motivacion problemas}
\end{figure}

As mentioned previously, two types of saturation will be studied in the setpoint-tracking problem: during the transient period and at steady state. Figure \ref{fig: Setpoint tracking motivacion problemas} shows an example of the process
\begin{equation}
    P(s)= \frac{1}{6s+1}e^{-s},
\end{equation}
with a PI controller tuned using $\lambda = 0.2T$, in which three reference changes are applied. The strategies shown in this example are DBC1, IBC, CI, DBC\_STr, H1 and H2. In the first-step change, the control signal saturates at $u_{max} = 2.5$ during the transient period. In this case, performance is degraded due to saturation, but all strategies eventually reach the setpoint. However, when the controller attempts to reach the second step change, the control signal saturates at $u_{min} = -0.6$, preventing it from reaching the setpoint until a new change is applied. Therefore, the study of this problem will be split into these two cases. In the first one (saturation on the transient), all the shown strategies will be compared. However, in the second case (saturation at steady state), DBC\_STr will not be included due to its poor performance across all cases. This can be seen in the third step change in Figure \ref{fig: Setpoint tracking motivacion problemas}, where this strategy takes much longer time to exit saturation, and therefore the process output response is considerably slower than that of the remaining strategies.

\subsubsection{Saturation on transient period}

\begin{figure}[h!]
    \centering
    \includegraphics[width=\textwidth]{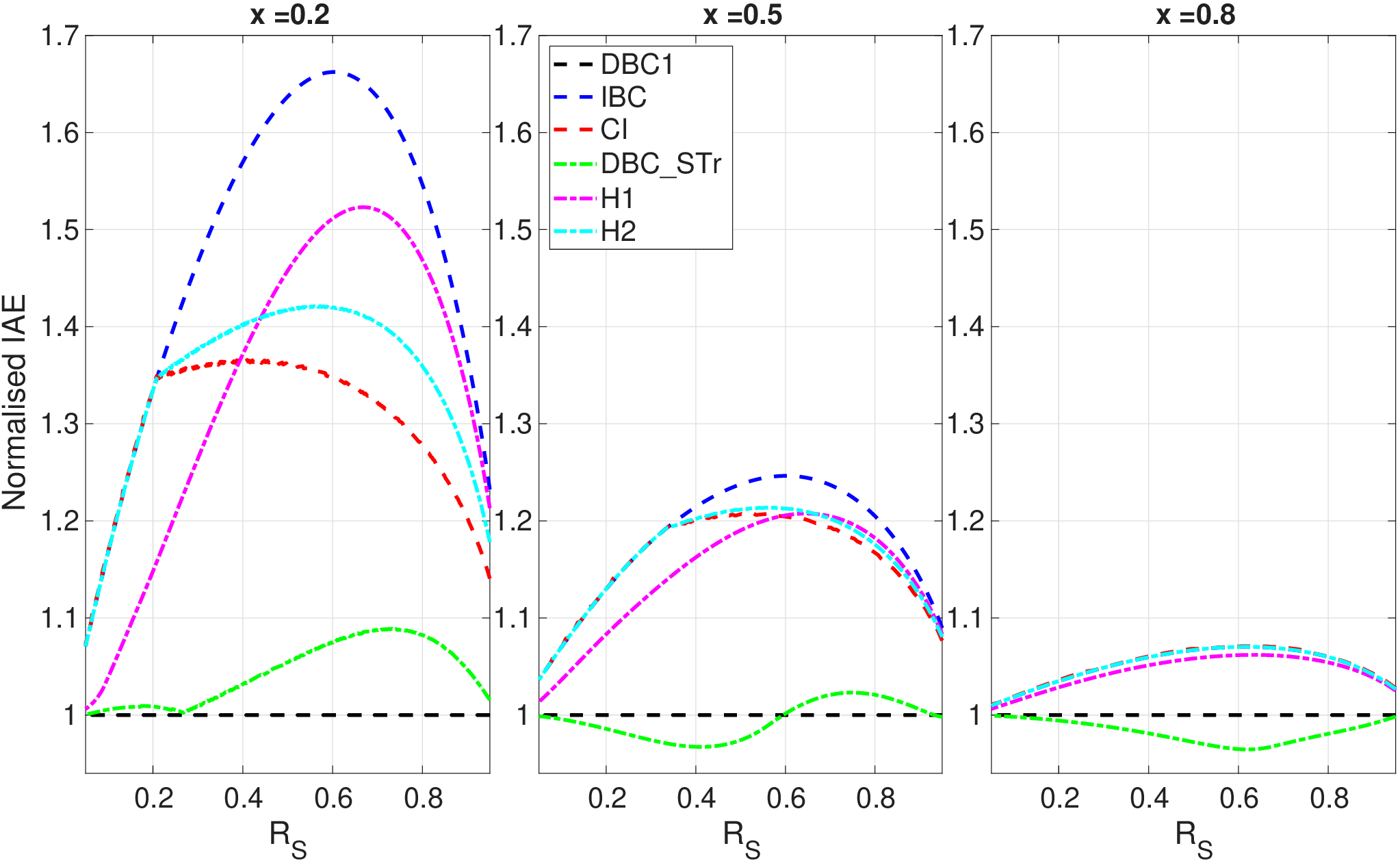}
    \caption{Performance comparison for processes with $L/T=1/6$}
    \label{fig: strategies comparison setpoint tracking lag lag}
\end{figure}

As mentioned before, for cases in which the transient of the control signal is saturated, the strategies compared are DBC1, IBC, CI, DBC\_STr, H1, and H2. Figure \ref{fig: strategies comparison setpoint tracking lag lag} shows the results obtained by each strategy for $T/L = 1/6$ and for each value of $x$ in Table \ref{tab: parametros barridos}. Each controller's aggressiveness is represented by a subfigure, in which the normalised IAE is plotted as a function of the saturation ratio. The performance metric was normalised to that obtained by DBC1, the most classical and widely implemented strategy. 

It can be seen that dynamic back-calculation is always the best anti-windup scheme, with strategy DBC1 being the best option for $x=0.2$, regardless of the saturation ratio. Moreover, DBC\_STr achieves a similar behaviour, with an IAE less than 10\% higher for saturation ratios over 0.3. On the other hand, the remaining schemes (IBC, CI, H1, and H2) exhibit worse behaviour, with IAEs up to 65\% higher for IBC. As the controller becomes more conservative, the best strategy depends on the saturation ratio, though it is always a trade-off between DBC1 and DBC\_Str. In the case of $x=0.5$, below $R_S=0.6$, DBC\_STr is preferable, whilst for greater values, DBC1 achieves a better performance. As $x$ increases, DBC\_STr becomes the best strategy, regardless of the saturation ratio, as evidenced by its performance at $x=0.8$. Lastly, the more conservative the controller, the less improvement one may achieve when choosing among these strategies. 

\begin{figure}[h!]
    \centering
    \includegraphics[width=\textwidth]{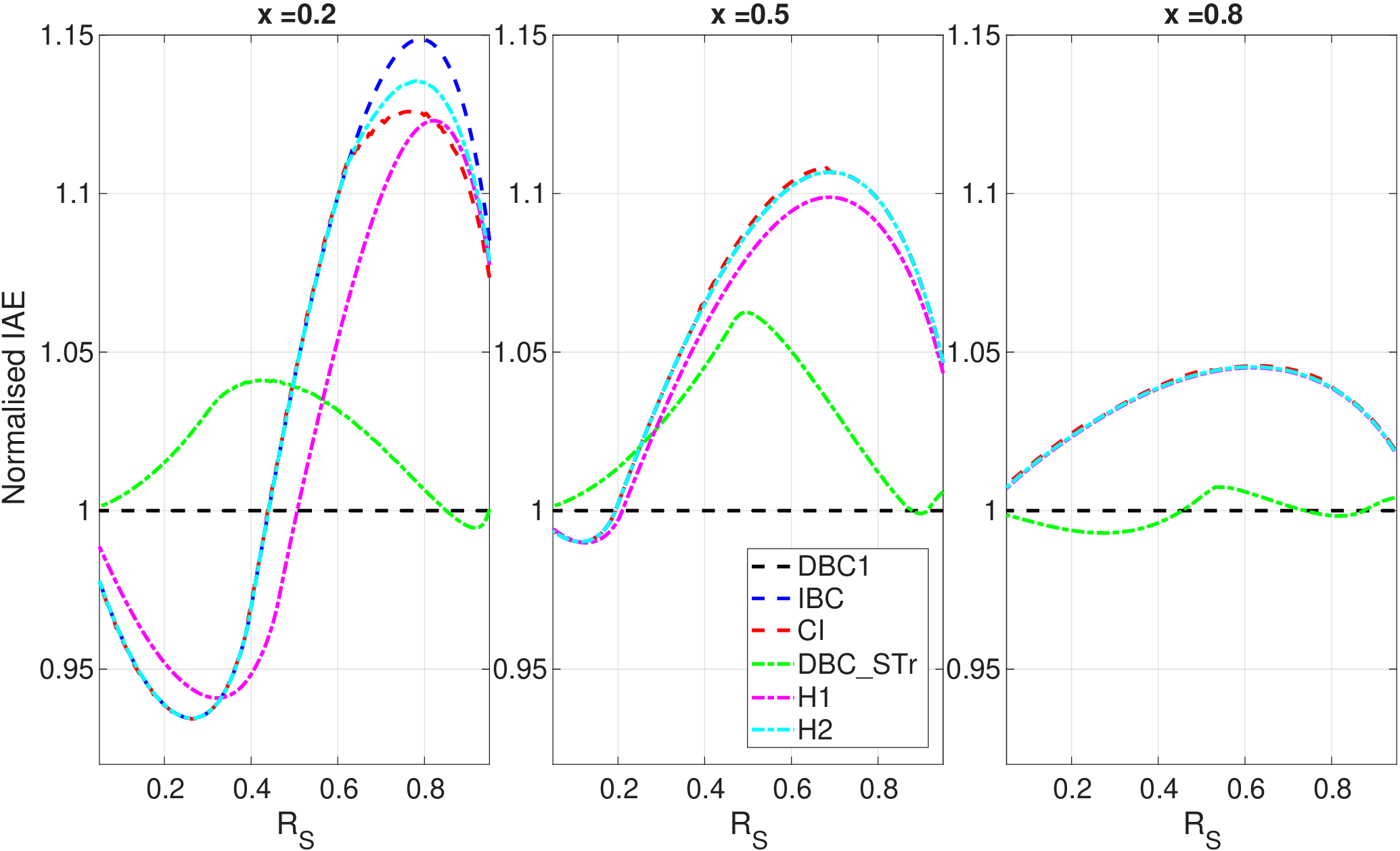}
    \caption{Performance comparison for processes with $L/T = 1/2$}
    \label{fig: strategies comparison setpoint tracking lag}
\end{figure}

For cases in which the delay is half the time constant, the results of the different strategies are shown in Figure \ref{fig: strategies comparison setpoint tracking lag}. In this case, DBC\_STr is never the best option, as it does not achieve the lowest IAE, except for a minimal improvement at $x = 0.8$. The optimal strategy varies among the remaining ones: for $x = 0.2$, if $R_S \leq 0.5$, IBC, CI, H1, and H2 can achieve up to an 8\% improvement. For higher saturation ratios, DBC1 obtains the best performance, being up to 15\% better than the other options. If $x$ is increased to $0.5$, the profile of the optimal strategies is maintained, but the saturation ratio lowers to $0.2$. However, just a 10 to 12\% improvement may be achieved in the best-case scenario. Lastly, for conservative controller tuning ($x = 0.8$), all strategies exhibit similar behaviour, although DBC1's IAE is slightly lower across almost the entire range of $R_S$. As happened in the case of processes with $L/T = 1/6$, the higher the value of $x$, the smaller the improvement among strategies. 

\begin{figure}[h!]
    \centering
    \includegraphics[width=0.6\linewidth]{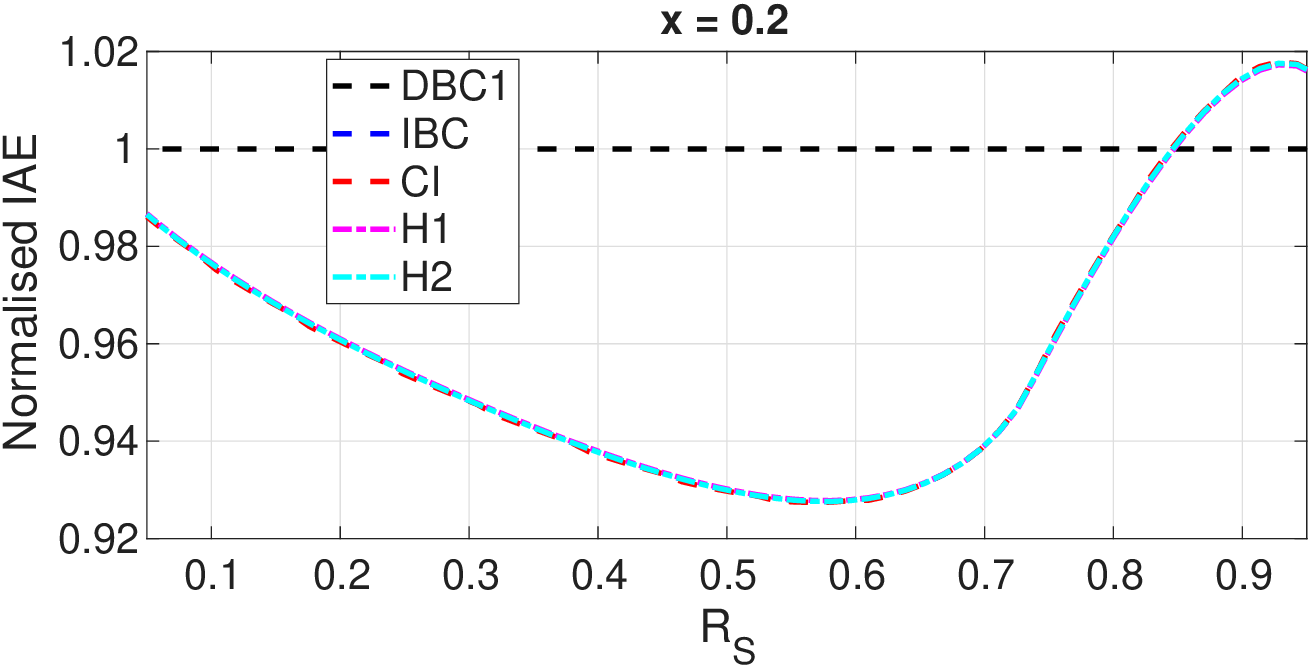}
    \caption{Performance comparison for processes with $L/T = 1$}
    \label{fig: strategies comparison setpoint tracking bal}
\end{figure}

Lastly, for processes in which $L = T$, all strategies are better than DBC1, obtaining an improvement of up to 7\% in performance for $x=0.2$ (see Figure \ref{fig: strategies comparison setpoint tracking bal}). It must be taken into account that DBC\_STr cannot be implemented for values of $L/T$ higher than $0.92$, due to the formulation of $\beta$ (see Equation \eqref{eq: Tt new}, which must be positive in order to obtain $T_t^{new} = \beta T_i$. 

\begin{figure}[ht!]
    \centering
        \centering
        \includegraphics[width=0.5\textwidth]{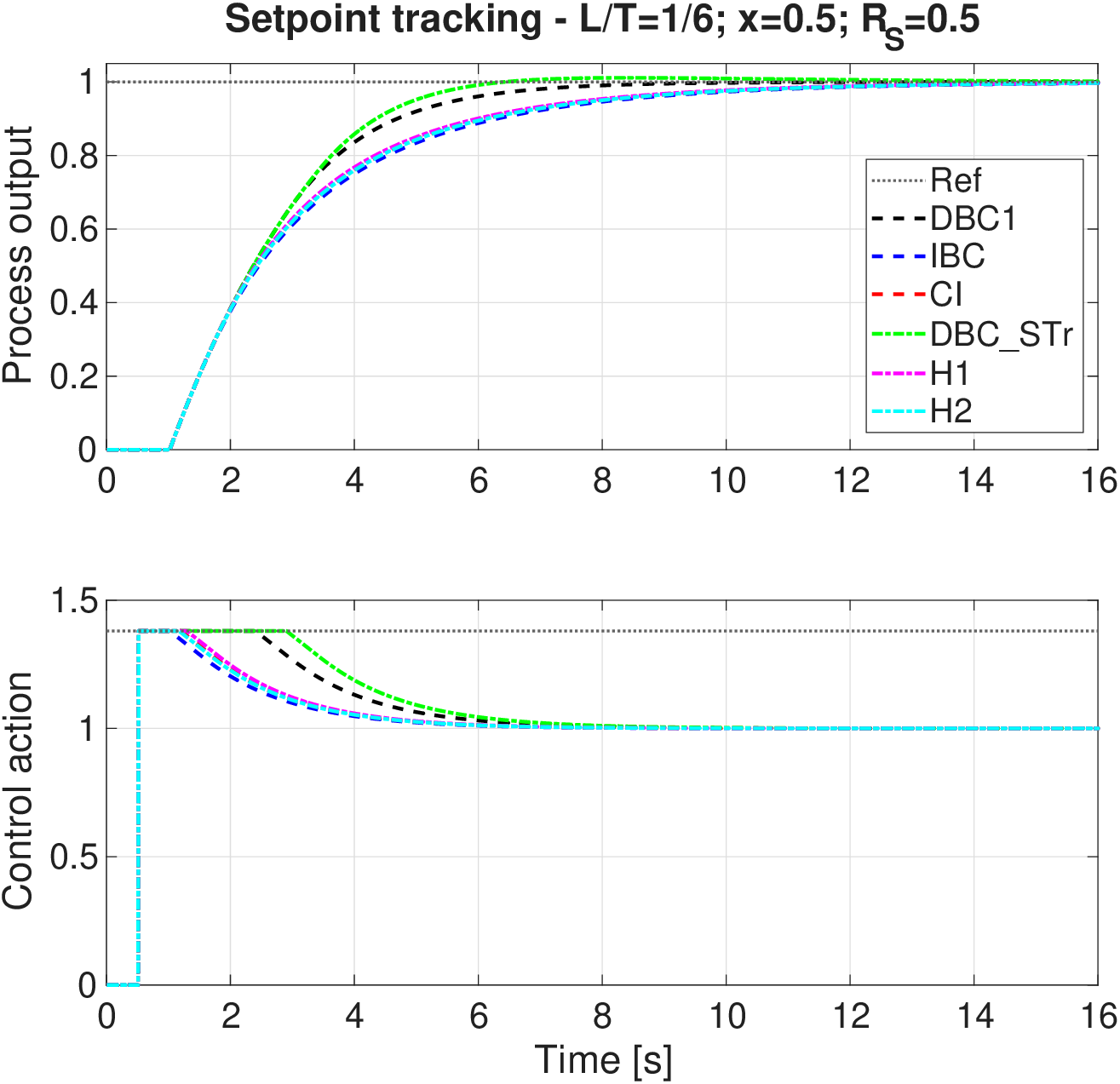}        
        \caption{Performance comparison for setpoint tracking - Temporal responses - 1}
        \label{fig:Ex11}
    \hfill
\end{figure}

\begin{figure}[ht!]  
\centering
    \begin{subfigure}[b]{0.4825\textwidth}
        \centering
        \includegraphics[width=\textwidth]{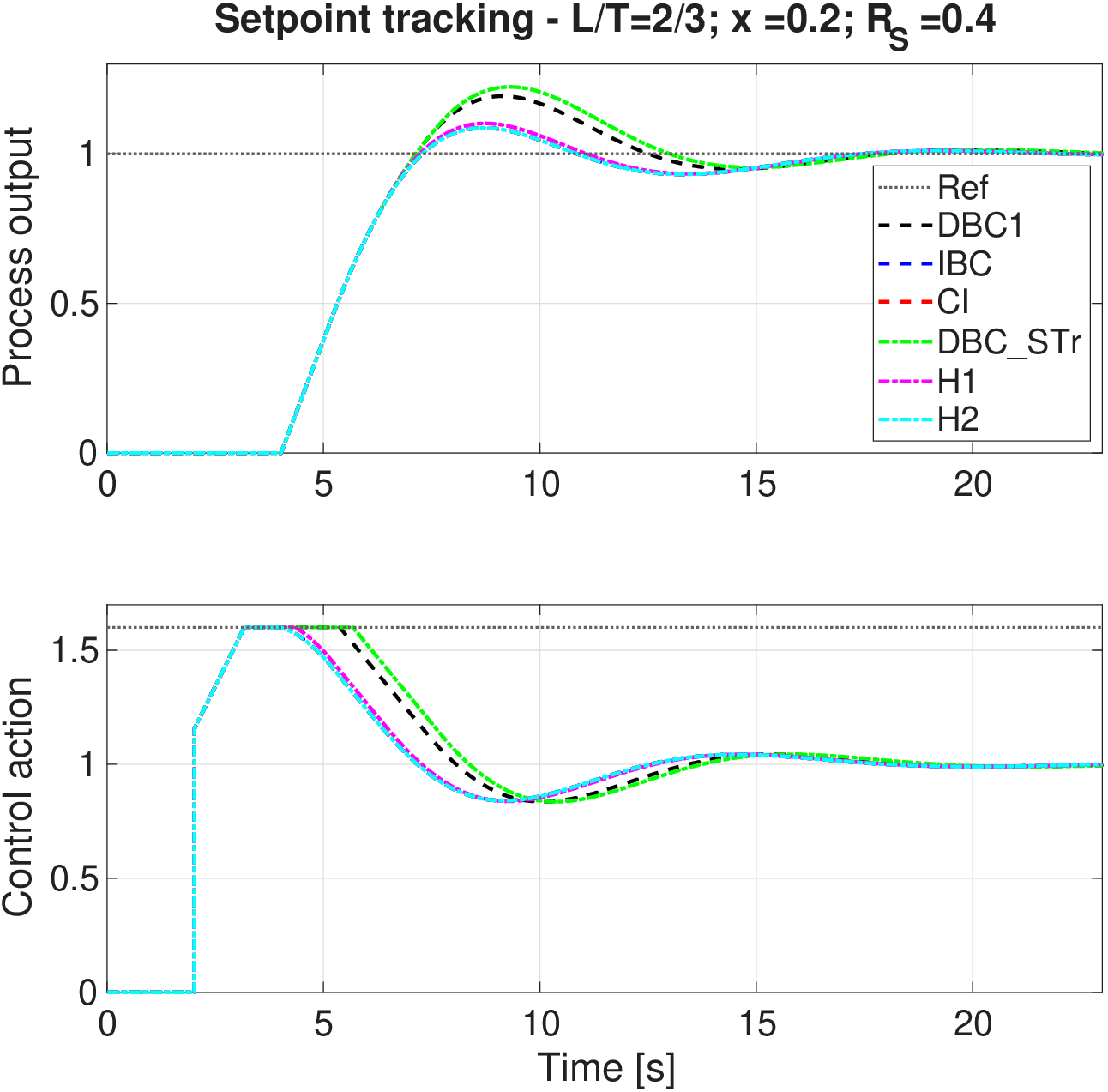}
        \caption{}
        \label{fig:Ex12}
    \end{subfigure}
    \hfill
    \begin{subfigure}[b]{0.49\textwidth}
        \centering
        \includegraphics[width=\textwidth]{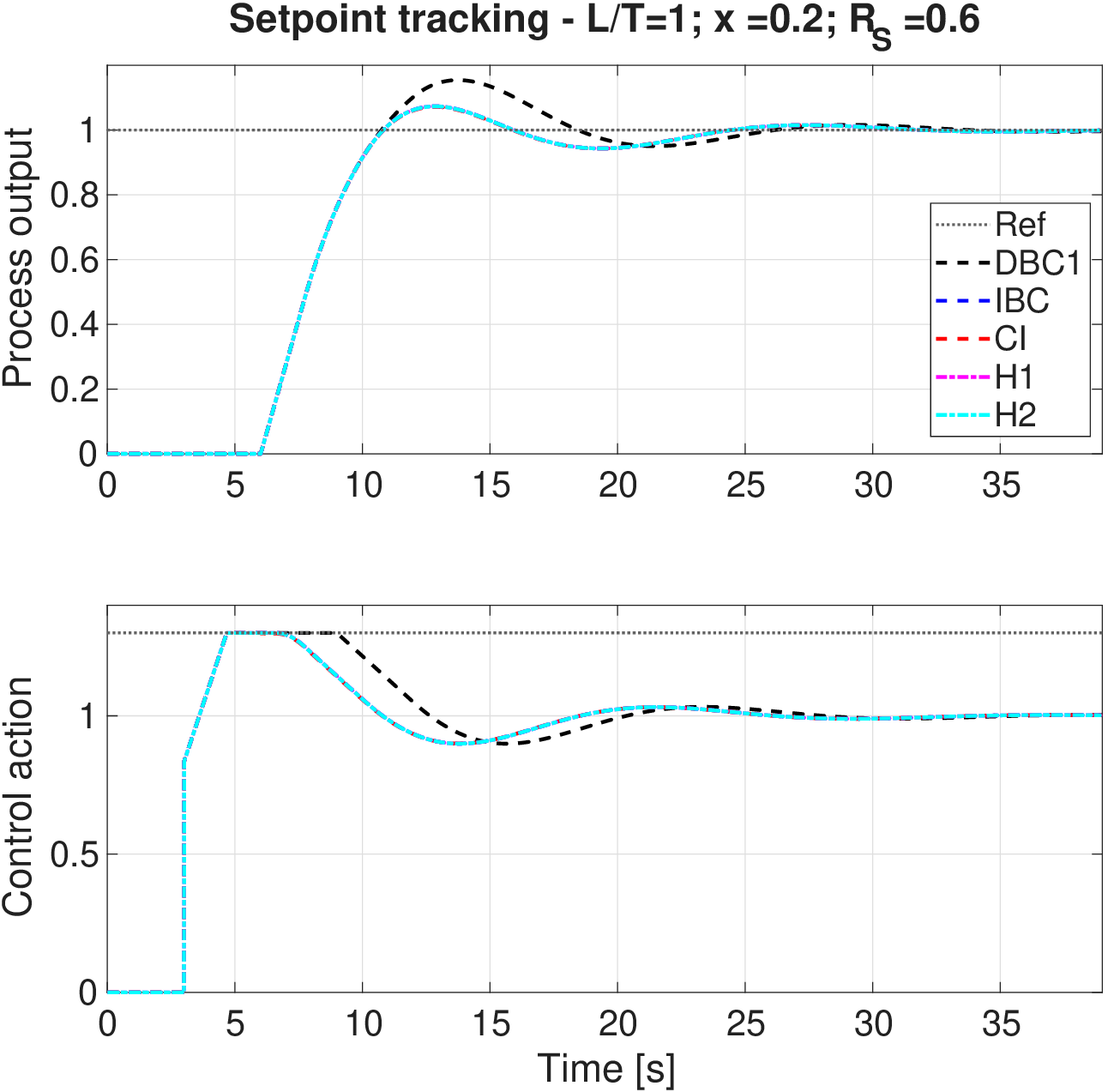}
        \caption{}
        \label{fig:Ex13}
    \end{subfigure}
    \caption{Performance comparison for setpoint tracking - Temporal responses - 2}
    \label{fig: strategies comparison setpoint tracking - temporal}
\end{figure}

After the extensive study of the different strategies' performance under different conditions, specific examples of the temporal response of each anti-windup strategy are shown in Figures \ref{fig:Ex11} and \ref{fig: strategies comparison setpoint tracking - temporal}, in which the studied process is 
\begin{equation}
    P(s) = \frac{1}{3s+1}e^{-Ls},
\end{equation}
for different values of delay, $L$. Figure \ref{fig:Ex11} shows an example for a lag-dominant process with a medium aggressiveness and a medium saturation ratio (0.5 for both parameters), in which both dynamic back-calculations achieve the fastest response, with DBC\_STr being slightly better. The remaining strategies, on the other hand, exhibit a slower response because they exit saturation more quickly. As the delay-time constant ratio increases, dynamic back-calculation degrades performance, resulting in greater overshoot in the process output. This behaviour is accentuated as the controller's aggressiveness is increased, as is the case of Figures \ref{fig:Ex12} and \ref{fig:Ex13}, in which strategies IBC, CI, H1, and H2 achieve the best performances, which are almost the same for all of them in both cases ($L/T = 2/3$ and $L/T=1$).

Therefore, the following guidelines may help the user choose the best strategy for setpoint-tracking problems in which saturation occurs during the transient period. If the process is highly lag-dominant ($L/T=1/6$), DBC1 is the best strategy for aggressive controllers ($x=0.2$), whilst DBC\_STr achieves slightly better performance for conservative tuning ($x=0.8$). For medium controller tuning, DBC\_STr performs better at low saturation ratios, whilst DBC1 is preferable at higher saturation ratios. Moreover, for this type of process, the more aggressive the controller, the worse the performance of the remaining strategies (IBC, CI, H1, and H2). For processes with medium delay-to-time-constant ratios ($L/T=1/2$) and aggressive controller tuning, IBC, CI, H1, and H2 perform better than the remaining strategies at low saturation ratios. In contrast, for high saturation ratios, DBC1 is the best alternative. For medium-aggressive tuning, DBC1 achieves the smallest IAE across almost the entire range of saturation ratios. For conservative tuning, both dynamic back-calculations (DBC1 and DBC\_STr) exhibit similar behaviour. Lastly, for balanced processes, IBC, CI, H1 and H2 are preferable under all possible conditions. \textcolor{black}{ Despite the fact that performance improvement may be achieved if the correct anti-windup technique is chosen, the user should pay special attention if their process is lag-dominant with medium-to-aggressive controller tuning, in which up to 70\% improvement may be observed. In the rest of scenarios, on the contrary, choosing the wrong technique may degrade the system's performance by 15\%, in the worst case.}

\subsubsection{Saturation on the steady state}

In the second type of saturation one may encounter in the setpoint-tracking problem, the reference is unreachable because the actuator's limit is smaller than the final control action required to achieve the goal. As mentioned previously, DBC\_STr is not included in this analysis because the first large $T_t$ allows the actuator to remain in saturation for a longer period, resulting in the worst performance across all studied conditions, with a very slow recovery of the process output. 

\begin{figure}[h!]
    \centering
    \includegraphics[width=\textwidth]{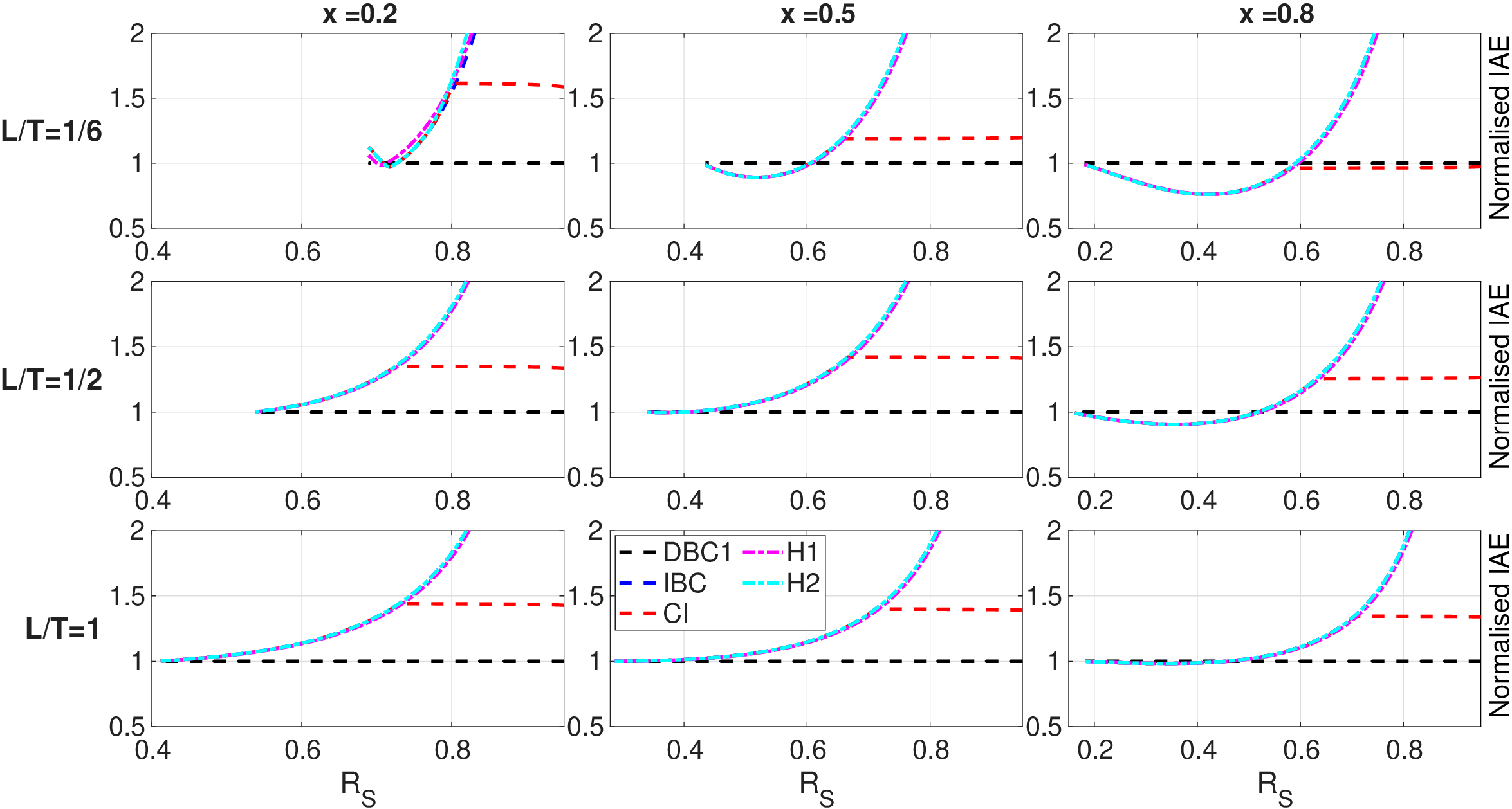}
    \caption{Performance comparison for unreachable setpoint problem}
    \label{fig:strategies comparison SP tracking saturation steadystate}
\end{figure}

Figure \ref{fig:strategies comparison SP tracking saturation steadystate} shows the normalised IAE achieved by the strategies, depending on the controller's aggressiveness ($x$), the delay dominance ($L/T$) and the saturation ratio ($R_S$). It can be seen that across all possible combinations, strategies IBC, H1, and H2 achieve the same IAE performance. Moreover, CI exhibits the same behaviour at lower saturation ratios. DBC1 is always the best alternative for aggressive controllers ($x=0.2$), medium-aggressive controllers ($x=0.5$), and medium-to-high delay dominance ($L/T \geq 1/2$). For cases with $x=0.5$ and $ L/T=1/6$, the remaining strategies are preferable for saturation ratios below $0.5$, \textcolor{black}{obtaining up to 10\% improvement}. For cases with conservative tuning, typically implemented in setpoint-tracking problems, at medium-to-low values of $R_S$, the best anti-windup schemes are IBC, CI, H1, and H2. For medium-to-high saturation ratios, DBC1 is the best strategy; CI is preferred for lag-dominant processes.

In general, DBC1 is the best strategy for setpoint-tracking problems in which the setpoint is unreachable, \textcolor{black}{except for lag-dominant processes with medium-to-conservative controller tuning, in which CI yields the smallest IAE across the entire range of $R_S$, with up to 25\% improvement. It should be noted that, choosing the wrong strategy in this case could result in a performance up to 15 times worse than the optimal, in cases with high saturation ratio. However, in order to ease the analysis of Figure \ref{fig:strategies comparison SP tracking saturation steadystate}, the y axis does not include these high values, as the IAE of strategies IBC, H1 and H2 increase exponentially.}

\begin{figure}[h!]  
\centering
    \begin{subfigure}[b]{0.48\textwidth}
        \centering
        \includegraphics[width=\textwidth]{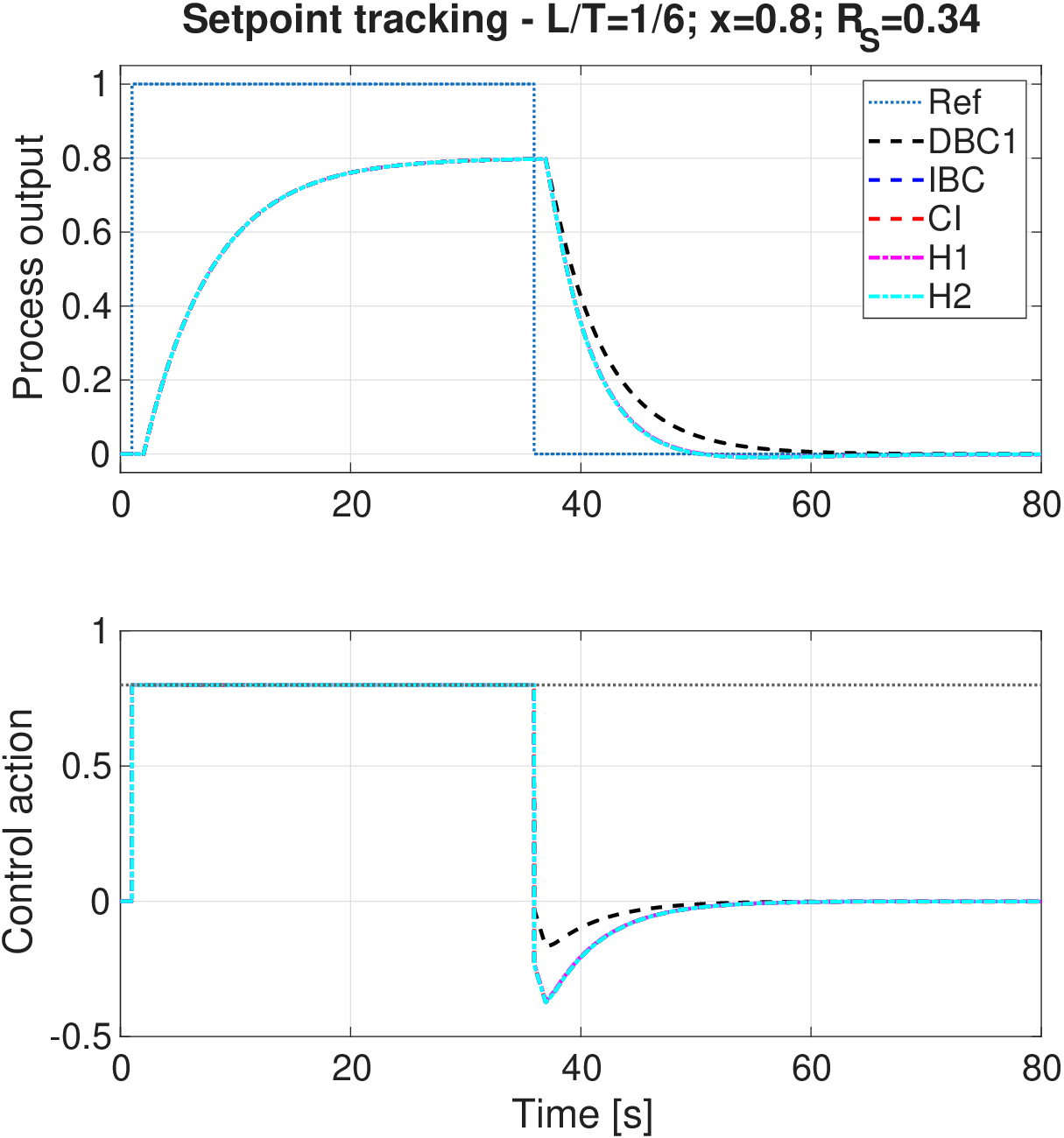}
        \caption{}
        \label{fig:Ex111}
    \end{subfigure}
    \hfill
    \begin{subfigure}[b]{0.48\textwidth}
        \centering
        \includegraphics[width=\textwidth]{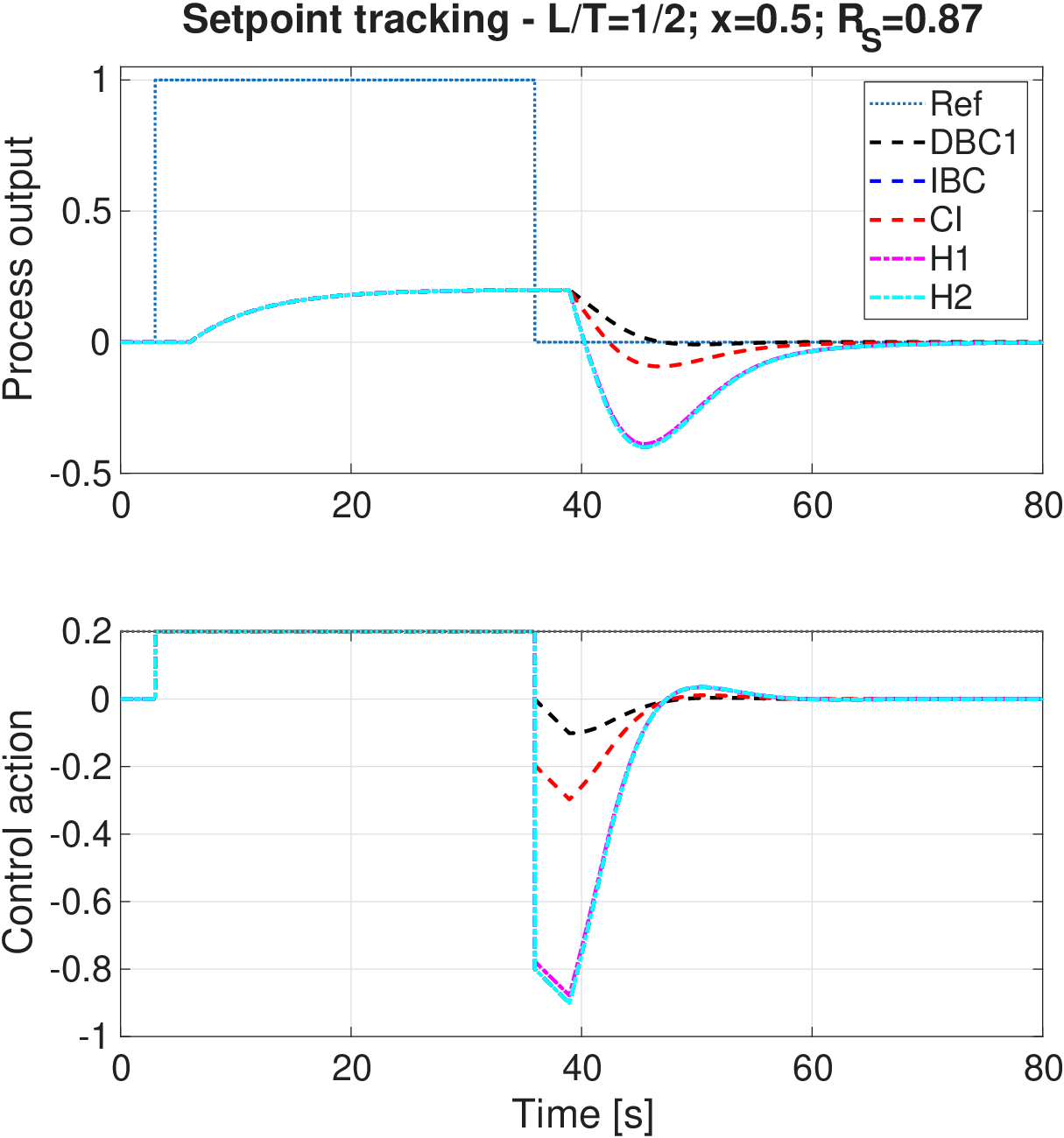}
        \caption{}
        \label{fig:Ex112}
    \end{subfigure}
    \caption{Performance comparison for setpoint tracking - Temporal responses}
    \label{fig: strategies comparison SP tracking 2 - temporal response}
\end{figure}

Figure \ref{fig: strategies comparison SP tracking 2 - temporal response} shows two examples of the process output response, in which both examples have $K = 1$ and $T=6$. Figure \ref{fig:Ex111} shows a very lag-dominant case, with $L/T=1/6$, with a conservative tuning ($x=0.8$) and a low saturation ratio. It can be seen that DBC1 obtains the slowest response, whilst the remaining strategies achieve a faster response thanks to a more aggressive change in the control action when the setpoint step is applied.

Figure \ref{fig:Ex112}, on the other hand, shows an example for $L/T=1/2$, with $x=0.5$ and a higher saturation ratio ($R_S = 0.87$). In this case, DBC1 achieves the best tracking with almost no overshoot on the process output. CI is the second-best strategy, but a clear overshoot is seen. Lastly, the aggressive behaviour of the remaining strategies yields the worst performance, with the largest overshoot in the process output.

\subsection{Disturbance-rejection problem}

\begin{figure}[h!]
    \centering
    \includegraphics[width=\textwidth]{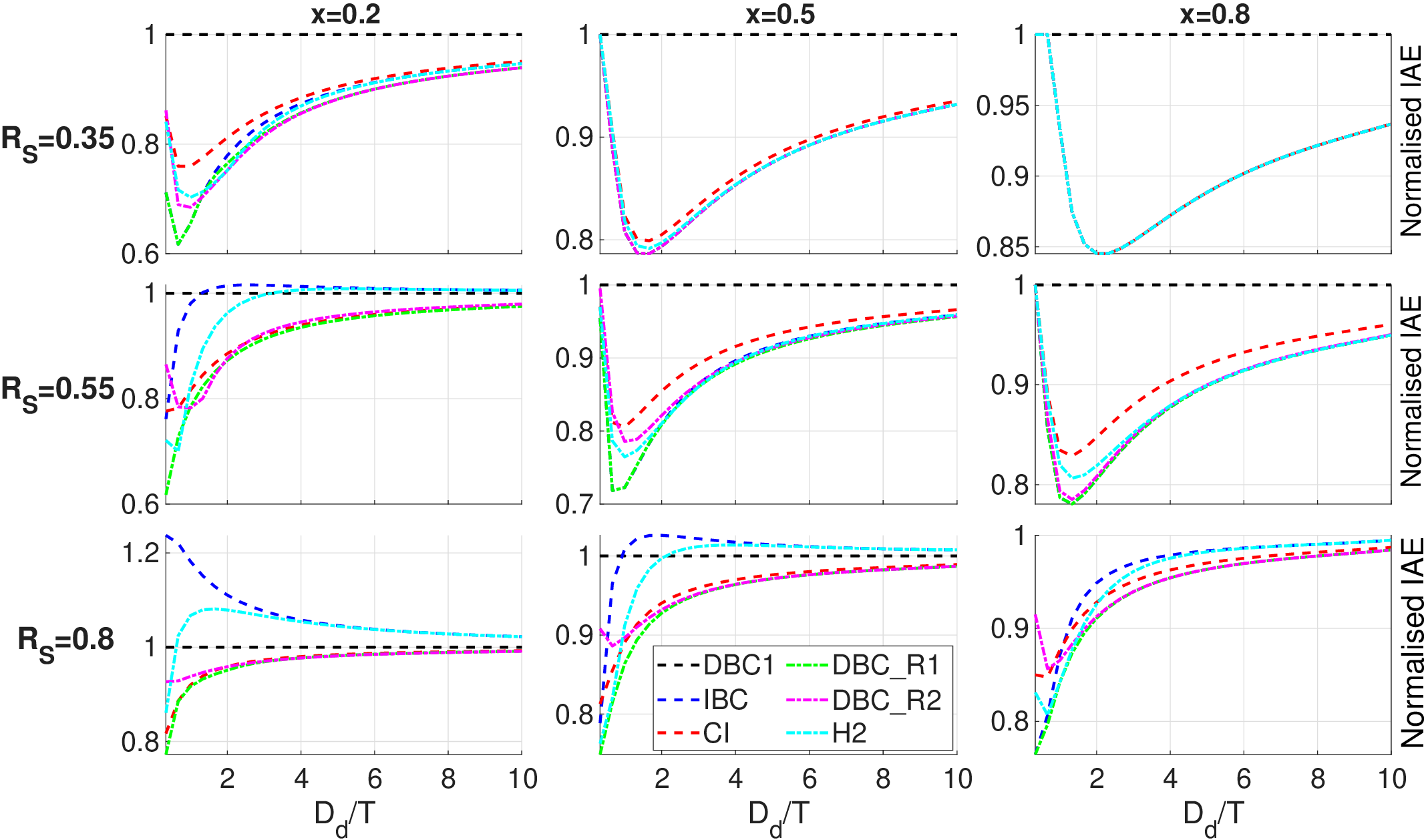}
    \caption{Performance comparison for processes with $L/T = 1/6$}
    \label{fig:strategies comparison no FF lag lag}
\end{figure}

\begin{figure}[h!]
    \centering
    \includegraphics[width=\textwidth]{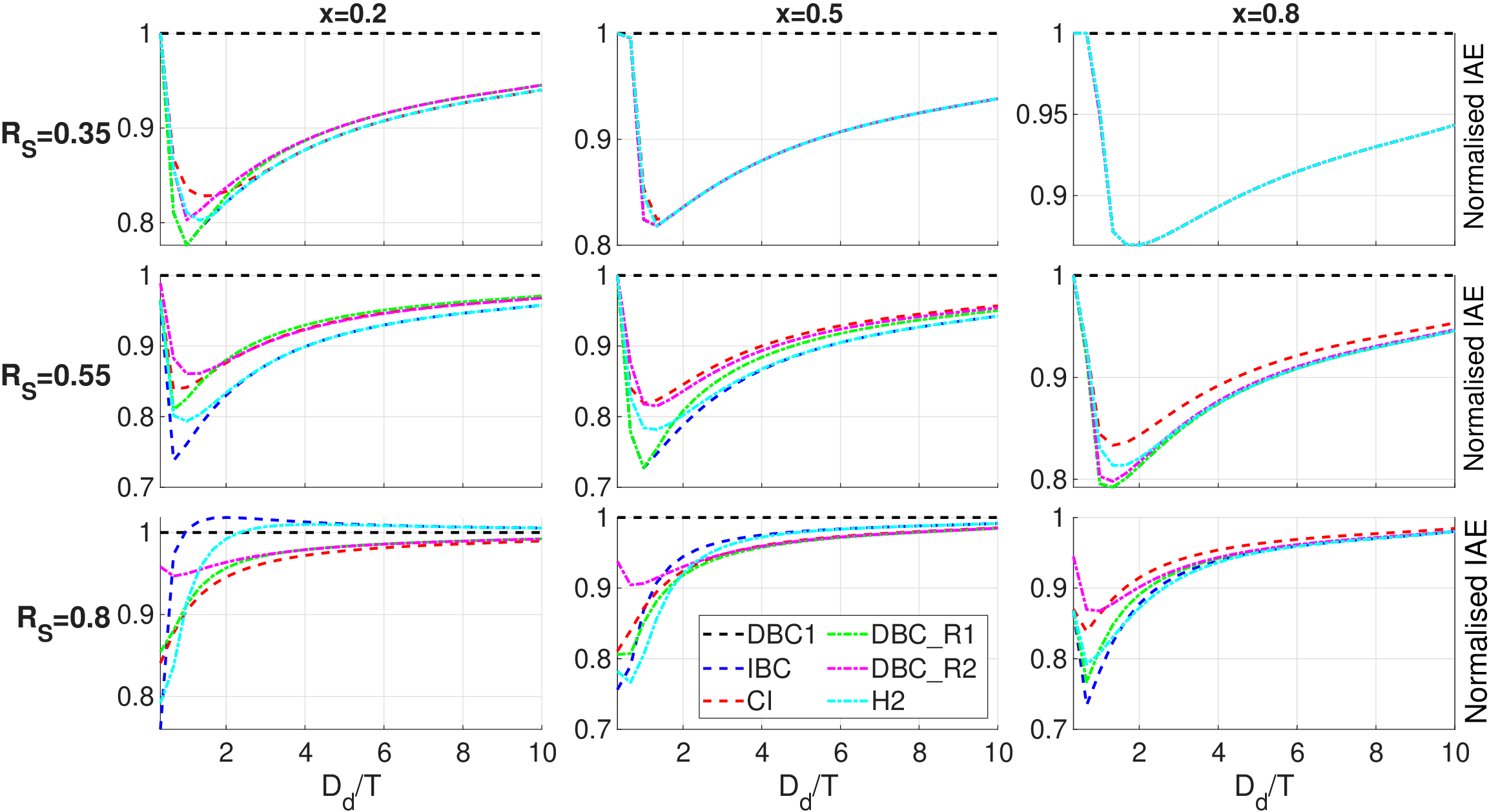}
    \caption{Performance comparison for processes with $L/T = 1/2$}
    \label{fig:strategies comparison no FF lag}
\end{figure}

\begin{figure}[h!]
    \centering
    \includegraphics[width=\textwidth]{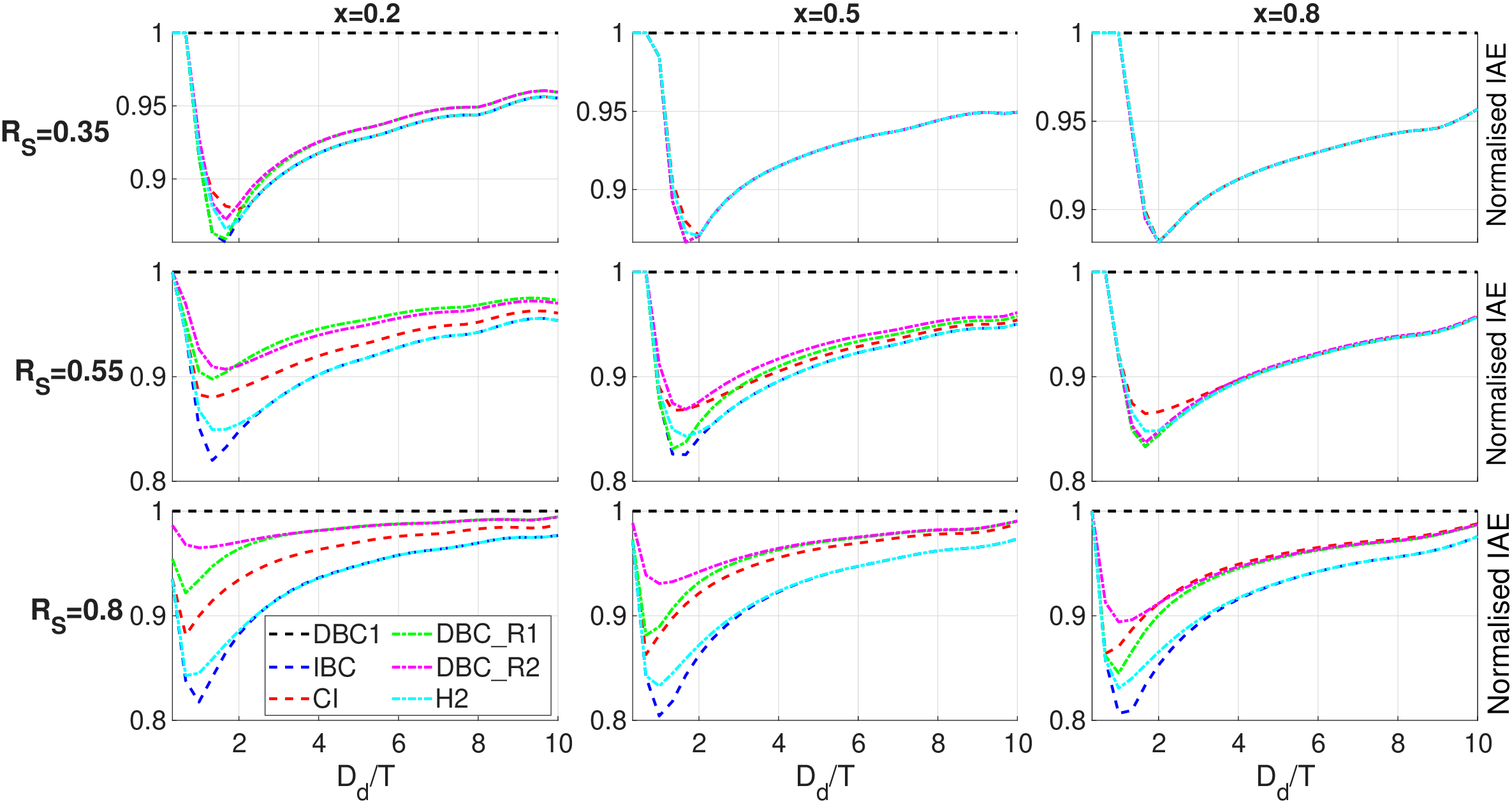}
    \caption{Performance comparison for processes with $L/T = 1$}
    \label{fig:strategies comparison no FF bal}
\end{figure}

In the disturbance-rejection problem, the strategies compared are DBC1, IBC, CI, DBC\_R1, DBC\_R2, and H2. Figures \ref{fig:strategies comparison no FF lag lag} to \ref{fig:strategies comparison no FF bal} show the results obtained by each strategy for different types of processes, where each row of figures represents a saturation ratio and each column a different aggressiveness. For each combination of these parameters, the IAE of each strategy is represented, normalised by that of DBC1, as a function of $D_d/T$. 

In the case of lag-dominant processes with low $L/T$ ratios (Figure \ref{fig:strategies comparison no FF lag lag}), it can be seen that, regardless of the controller's aggressiveness, for low-to-moderate saturation ratios, DBC1 is always the worst option and considerable performance improvement may be achieved by implementing other strategies. 

For aggressive controller tuning ($x = 0.2$), DBC\_R1 achieves the best behaviour across the entire range of disturbance durations and all studied saturation ratios. DBC\_R2 obtains the same behaviour for $D_d/T >1$, but presents a worse behaviour for shorter disturbances. IBC and H1 achieve close-to-optimal performance at low saturation ratios, but at medium-to-high $R_S$ values, they perform poorly, often worse than DBC1. Lastly, CI performs, on average, better than IBC and H1, but in some cases the obtained IAE is 15\% higher than the optimal value. 

For a medium-to-low-aggressiveness controller ($x = 0.5$ and $ x = 0.8$), the results are similar to those obtained with aggressive tuning and medium-to-low saturation ratios ($R_S \leq 0.5$). However, for low values of $R_S$, all strategies perform similarly, except for DBC1, which is up to 20\% worse.

For cases with medium delay-to-time-constant ratios, as is the case of Figure \ref{fig:strategies comparison no FF lag}, where $L/T=1/2$, once again DBC1 is never the best option available. For low saturation ratios, the remaining strategies perform almost the same for all values of $x$, with a small difference in the case of $x=0.2$, where DBC\_R1 obtains the best performance along with IBC.  

For intermediate values of $R_S$, IBC is the best option with the lowest IAE, followed by H2 and DBC\_R1. CI and DBC\_R2 exhibit similar behaviour, but they yield IAEs up to 10\% higher than the optimal value, especially for $D_d/T$ close to 1. These characteristics may also be observed at high saturation ratios and with medium-to-conservative tuning. However, for $x = 0.2$ and $R_S = 0.8$, the pattern changes, as IBC turns into one of the worst options. H2 presents the optimal behaviour for $D_d/T \leq 1$, whilst achieving the same performance as IBC for higher values of $ D_d/T$. The best performance, on average, for all disturbance lengths, is achieved by CI and DBC\_R1, along with DBC\_R2 for disturbances that last longer than $2T$. 

Lastly, balanced processes with $T=L$ were studied, obtaining the results shown in Figure \ref{fig:strategies comparison no FF bal}. For low saturation ratios, all strategies except DBC1 exhibit similar behaviour, with performance up to 15\% better than DBC1. Regardless of the controller's aggressiveness, as the saturation ratio increases, IBC and H2 obtain the best performance, improving by up to 20\%. In this process, although they outperform DBC1, the strategies DBC\_R1, DBC\_R2, and CI are not among the best options for achieving high $R_S$ values.

\begin{figure}[h!]  
\centering
    \begin{subfigure}[b]{0.49\textwidth}
        \centering
        \includegraphics[width=\textwidth]{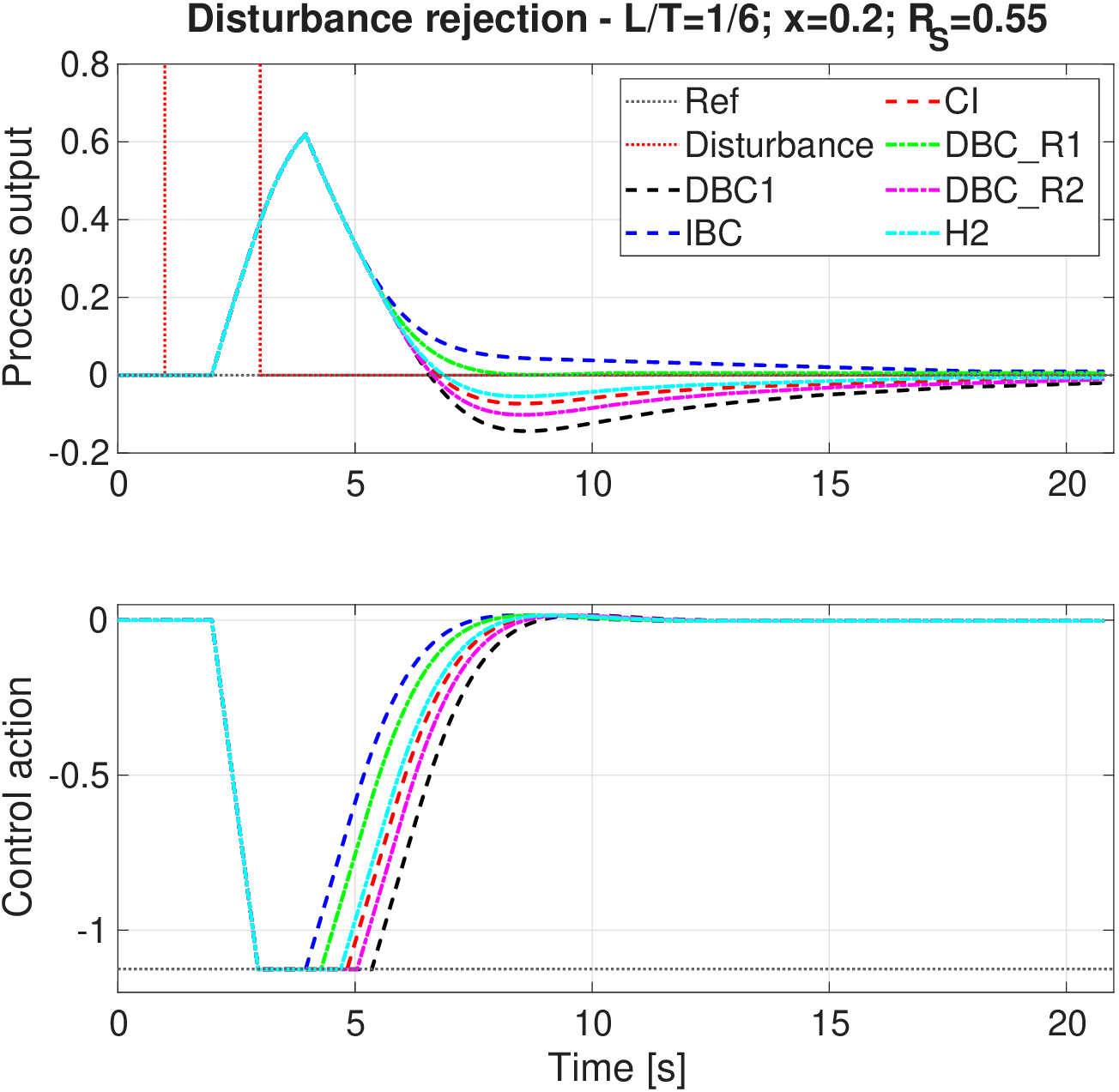}
        \caption{}
        \label{fig:Ex14}
    \end{subfigure}
    \hfill
    \begin{subfigure}[b]{0.49\textwidth}
        \centering
        \includegraphics[width=\textwidth]{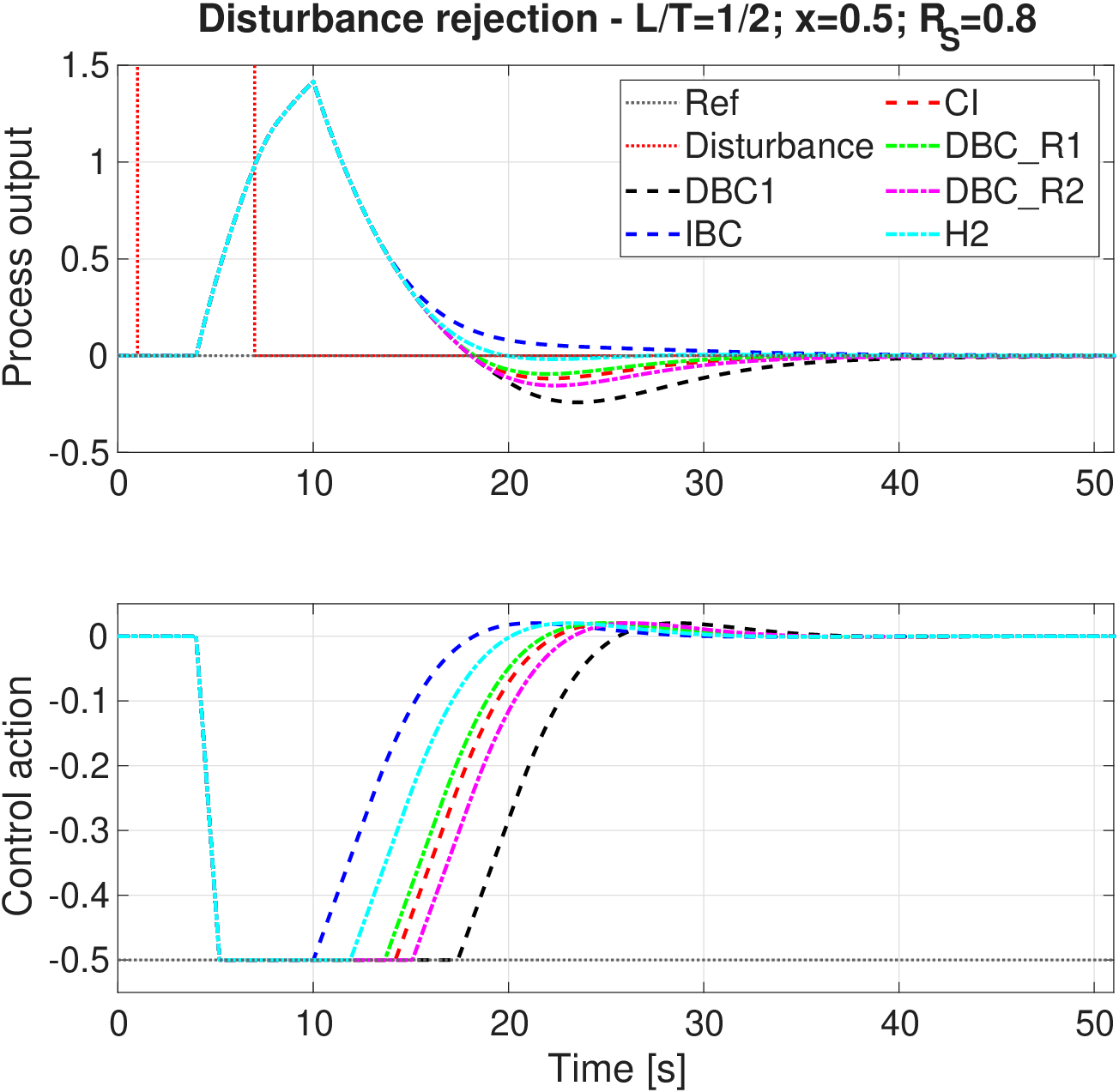}
        \caption{}
        \label{fig:Ex15}
    \end{subfigure}
    \hfill
    \begin{subfigure}[b]{0.49\textwidth}
        \centering
        \includegraphics[width=\textwidth]{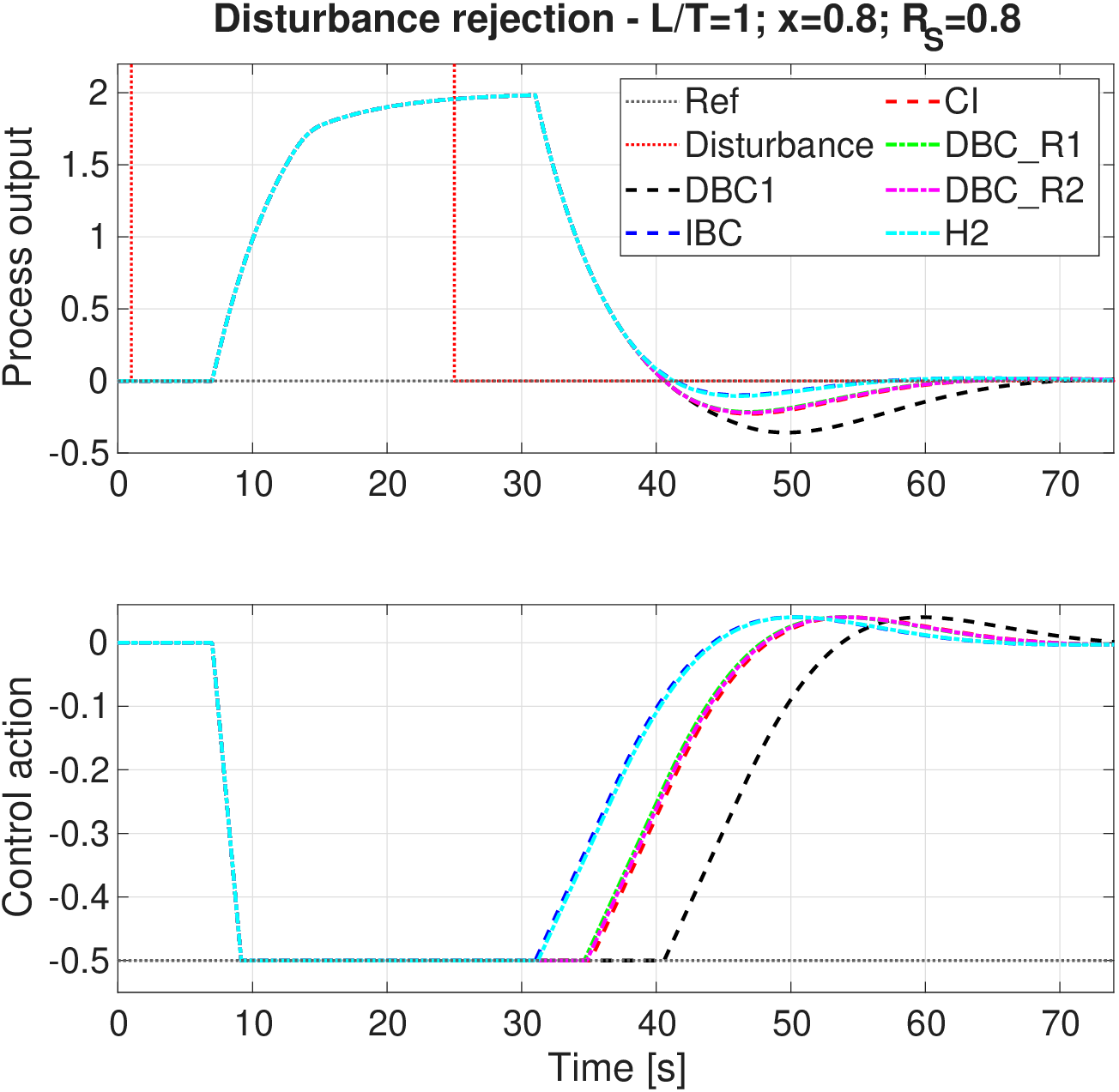}
        \caption{}
        \label{fig:Ex16}
    \end{subfigure}
    \caption{Performance comparison for disturbance rejection - Temporal responses. The first subplot shows the process output, together with the reference (in dotted black) and the disturbance (in dotted red), whilst the second subplot shows the applied control action for each strategy. }
    \label{fig: strategies comparison no FF - temporal}
\end{figure}

Once the general behaviour of the selected strategies was analysed and compared, some specific examples were examined, whose temporal responses are shown in Figure \ref{fig: strategies comparison no FF - temporal} for different configurations. 

For $L/T = 1/6$, $x=0.2$, $R_S = 0.55$ and $D_d/T = 1/3$ (Figure \ref{fig:Ex14}), DBC\_R1 is the best strategy, obtaining the fastest disturbance rejection with no overshoot, obtaining an IAE 38\% lower than DBC1, which obtains the largest overshoot in the process output. H2 is the second-best strategy, with a slight overshoot and a slower recovery, followed by IBC, which obtains a slow response without overshoot.

For a process with $L/T=1/2$, $x = 0.5$, $R_S = 0.8$, and $D_d = T$ (Figure \ref{fig:Ex15}), H2 and DBC\_R1 switch roles: H2 positions as the best strategy, recovering faster and with almost no overshoot, whilst DBC\_R1 takes longer to reach the reference and presents a small overshoot in the process output. Once again, IBC is slow but without overshoot, while DBC1 exhibits the worst response with the largest overshoot. Lastly, for a balanced process with conservative tuning ($x=0.8$), $R_S = 0.8$, and $D_d/T>4$ (Figure \ref{fig:Ex16}, IBC and H2 are the best anti-windup strategies, followed by CI, DBC\_R1 and DBC\_R2. However, in this case, all schemes reject the disturbance with an overshoot in the measured output.

\section{Conclusions}       \label{sec: conclusions}

This paper has presented an extensive analysis of anti-windup strategies for PI-controlled systems subject to actuator saturation, with a strong focus on practical applicability and performance optimisation. The results highlight that the choice and tuning of the anti-windup scheme play a critical role in determining closed-loop performance, and that commonly used rules of thumb are often far from optimal.

Classical techniques such as dynamic back-calculation and conditional integration remain competitive due to their simplicity, but their effectiveness strongly depends on process dynamics, controller aggressiveness, and the nature of saturation. Instantaneous back-calculation and conditional schemes tend to perform better in scenarios with strong delay dominance or low saturation levels, whilst dynamic back-calculation is generally more suitable for aggressive controllers and higher saturation ratios.

The proposed hybrid anti-windup strategy, which combines conditional integration with back-calculation, demonstrates improved responsiveness during saturation events and enhanced robustness against large disturbances, offering a practical compromise between simplicity and performance.

A major contribution of this work lies in the development of systematic tuning rules for the tracking time constant in back-calculation schemes. These rules, derived from a large-scale optimisation study, provide clear guidelines based on measurable system characteristics such as saturation ratio, controller aggressiveness, and disturbance dynamics. The results show that the optimal tracking time constant is generally smaller than the integral time and varies significantly with operating conditions, contradicting conventional heuristic recommendations.

The comparative study further shows that no single anti-windup strategy is universally optimal; instead, the best choice depends on the specific control objective (setpoint tracking vs. disturbance rejection), process characteristics, and available system information. However, the proposed tuning rules and selection guidelines enable practitioners to achieve near-optimal performance with minimal effort. Overall, this work bridges the gap between theoretical developments and industrial practice by providing intuitive, implementable, and performance-oriented solutions for anti-windup design and tuning. 

\section*{Acknowledgments}
This work has been financed by the following projects: PID2023-150739OB-I00 and PDC2025-165379-I00 financed by the Spanish Ministry of Science. The Lund Univesity co-authors are members of the ELLIIT strategic research area. Malena Caparroz acknowledges the financial support of the Spanish Ministry of Science, Innovation, and Universities under grant FPU23/02235.

\bibliographystyle{apacite}
\bibliography{Biblio}

\end{document}